# Nonlinear transport in a photo-induced superconductor

E. Wang[1], J. D. Adelinia[1], M. Chavez-Cervantes[1], T. Matsuyama[1], M. Fechner[1], M. Buzzi[1], G. Meier[1], A. Cavalleri[1,2]

[1] Max Planck Institute for the Structure and Dynamics of Matter, Hamburg, Germany

[2] Department of Physics, Clarendon Laboratory, University of Oxford, United Kingdom

**Optically driven quantum materials exhibit a variety of non-equilibrium functional phenomena [1-11], which are potentially associated with unique transport properties. However, these transient electrical responses have remained largely unexplored, primarily because of the challenges associated with integrating quantum materials into ultrafast electrical devices. Here, thin films of $K_3C_{60}$ grown by Molecular Beam Epitaxy were connected by coplanar terahertz waveguides to a series of photo-conductive switches. This geometry enabled ultrafast transport measurements at high current densities, providing new information on the photo-induced phase created in the high temperature metal by mid-infrared excitation [12-16]. Nonlinearities in the current-voltage charactersitics of the transient state validate the assignment of transient superconductivity, and point to an inhomogeneous phase in which superconducting regions of the sample are connected by resistive weak links [17-23]. This work opens up the possibility of systematic transport measurements in driven quantum materials, both to probe their properties and to integrate them into ultrafast optoelectronic platforms.**

Alkali-doped fullerites are strongly correlated organic superconductors that exhibit a high degree of tuneability. This is reflected in the phase diagram of Fig. 1b, which displays metallic and superconducting phases accessed by applying small amounts of physical or chemical pressure [24-29].

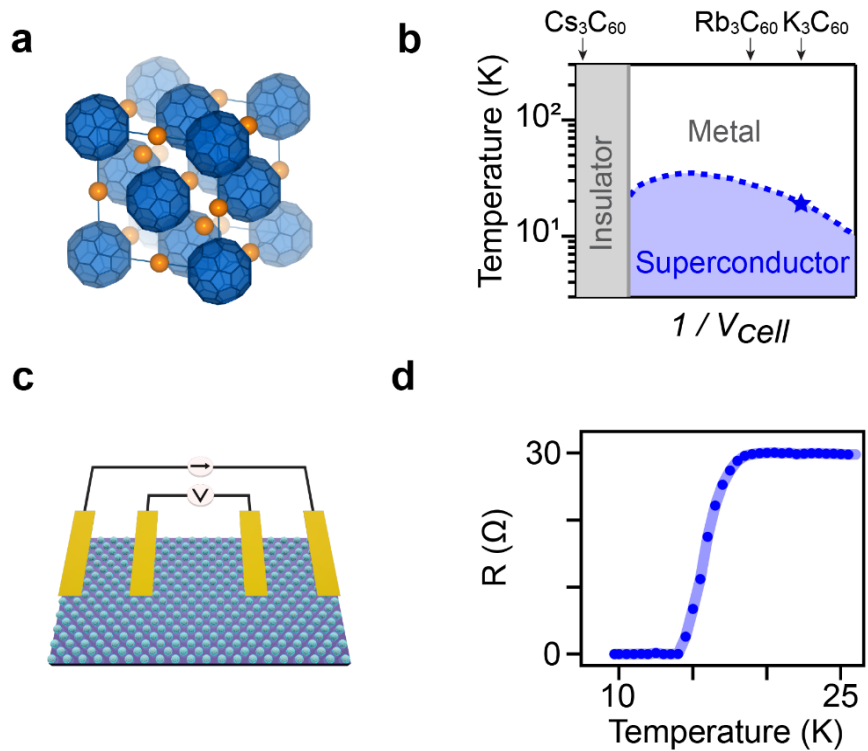

**Figure 1 | Crystal structure, phase diagram and four-point transport measurement of $K_3C_{60}$. a**, Crystal structure of the organic molecular superconductor $K_3C_{60}$. The molecules $C_{60}$ (blue) are arranged in the face-centered cubic structure with potassium atoms (orange) occupying interstitial voids. **b**, Phase diagram of the alkali-doped fullerene $A_3C_{60}$ family as function of the inverse of the unit cell volume $1/V_{cell}$. $V_{cell}$ can be tuned by chemical or physical pressure. The solid grey line indicates the boundary between insulating and metallic/superconducting phases. The blue dashed line shows the dependence of the superconducting transition temperature $T_c$ on $1/V_{cell}$. The star highlights the $K_3C_{60}$ compound investigated in this work, which becomes superconducting at $T < T_c = 19$ K. **c**, Schematic of a four-point DC transport measurement on a $K_3C_{60}$ thin film. **d**, Measured resistance versus temperature of the $K_3C_{60}$ thin film with four-point DC transport measurement.

A number of recent experiments have shown that these materials can also be manipulated with light, revealing signatures of optically induced superconductivity at temperatures far in excess of the equilibrium $T_c$. In $K_3C_{60}$, coherent optical driving of the high temperature metallic phase with intense mid infrared pulses ($\lambda \sim 7$ μm) has been shown to induce a transient state with superconducting-like optical properties [12, 13, 15]. This assignment was further supported by measurements performed under pressure, which showed a characteristic suppression of the non-equilibrium phase [13]. Electrical transport measurements showed a vanishingly small electrical resistance, also compatible with photo-induced superconductivity [15]. However, although all these observations are highly suggestive of a non-equilibrium superconducting state induced by light, important indicators like Meissner diamagnetism and critical current responses have not been reported yet. New experimental methods are needed to probe these properties at ultrafast timescales.

Here, a new ultrafast experimental platform was developed to probe linear and nonlinear electrical transport in photo-excited $K_3C_{60}$. Photo-conductive switches [30] and Terahertz (THz) waveguides were used to generate, transport and detect electrical pulses of only ~1 picosecond duration, which were confined to deeply sub-wavelength length scales in spectral ranges (50 GHz – 1 THz). This method accesses features that would be difficult to access with free space terahertz technology [31, 32].

Thin films of $K_3C_{60}$ were grown on sapphire substrates by molecular beam epitaxy, and optimized to yield a superconducting phase with $T_c$ = 19 K (details see Supplementary Section S2). A schematic of the geometry is shown in Fig. 2a. The $K_3C_{60}$ films were connected to three pairs of photo-conductive switches, consisting of amorphous silicon patches fabricated along the signal line of coplanar waveguides. When irradiated with 250 fs-long, 515 nm-wavelength laser pulses, the biased photo-conductive switches become transiently conductive, and launch ultrafast electrical pulses with duration determined by the lifetime of the photo-carriers in the silicon switches (~ 300 fs, see Fig. S8) and by the bandwidth of the waveguide (~1 THz, see Fig. S11). In all the measurements reported here, the two leftmost switches were excited simultaneously, to launch quasi-transverse electromagnetic (TEM) pulses along the signal line (details see Supplementary Section S3.4).

Before interaction with the $K_3C_{60}$ thin film, the electrical pulses were sampled in a second pair of switches (see Fig. 2a), providing a reference voltage transient $V_1(t)$. Unlike for the first pair of switches used to launch the pulses, this second pair of switches was left unbiased, yielding a measurable current only when *both* the electrical pulse *and* a femtosecond optical gate were superimposed on the switch. By scanning the mutual delay (denoted here as sampling time t) between the electrical pulse and the optical gating pulse, $V_1(t)$ could be measured (see representative transient in Fig. 2a). The pulse transmitted after interaction with the $K_3C_{60}$ sample was measured in the same way at a third pair of photo-conductive switches, yielding a second time-resolved trace $V_2(t)$, displayed in Fig. 2a.

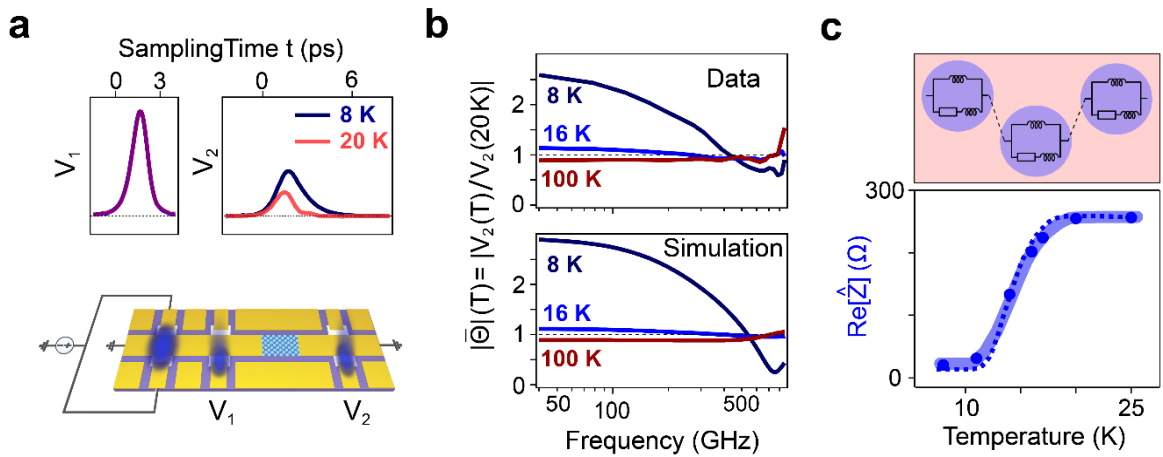

Figure 2| **Probing the equilibrium superconducting transition with ultrafast transport. a,** Lower: Schematic of ultrafast transport device. The MBE grown $K_3C_{60}$ thin film (cyan) and three pairs of Auston switches (white) were incorporated within a coplanar waveguide (yellow) on a sapphire substrate (purple). Ultrashort electrical pulses were launched by illuminating laser pulses onto the left pair of Auston switches, which were simultaneously biased by a voltage source. The launched pulse $V_1(t)$ was sampled by illuminating one Auston switch of the middle pair and by changing the mutual delay between the launching and sampling laser pulses. Similarly, the transmitted pulse $V_2(t)$ was sampled with one Auston switch of the right pair. Upper: $V_1(t)$ and $V_2(t)$ measured at 8 K and 20 K. Both measurements were normalized by the peak value of $V_1(t)$. **b,** Upper: Normalized transmittance $|\bar{\Theta}|$ measured at 8 K, 16 K and 100 K. Lower: Corresponding simulation using the effective circuit model shown in **c**. The black dashed lines indicate unity. **c,** Upper: Schematic of circuit model used in **b**. Blue circles indicate superconducting grains and red area indicates the resistive weak coupling between grains. Lower: Real part of the impedance $Re[\hat{Z}]$ at 50 GHz (blue dot) as function of temperature. The resistance versus temperature curve measured with DC transport on the same device is plotted as blue dashed line.

A few characteristics are immediately apparent by inspecting the $V_2(t)$ transients transmitted by the sample at equilibrium. For temperatures T > $T_c$, for which the sample was resistive, the voltage pulse was attenuated. For the low temperature superconductor (T < $T_c$), the transmitted $V_2(t)$ pulse was not attenuated, but was broadened in time. This effect is well understood by noting that superconducting $K_3C_{60}$ acts as lumped inductor, with no dissipation but with frequency dispersion ($\hat{Z} = j\omega L$). Note also that for T<$T_c$ the inductive response of the film causes a ~ 500 fs group delay relative to the resistive response, as expected.

The same data is presented as a frequency-resolved transmittance $\Theta(\omega, T) = V_2(\omega, T)/V_1(\omega, T)$ in Fig. 2b, plotted as $\overline{\Theta}(\omega, T) = \frac{\Theta(\omega,T)}{\Theta(\omega,20\ K)} = \frac{V_2(\omega,T)}{V_2(\omega,20\ K)}$ after normalization by the transmittance measured immediately above $T_c$. Note that the modulus of $\overline{\Theta}(\omega, T)$ was larger than 1 for all temperatures T < $T_c$ and smaller than 1 for all temperatures T > $T_c$ (more temperatures see Fig. S28). The dip in transmission near 700 GHz, observed only in the superconducting state, reflects an LC resonance formed by the inductance of the superconductor $L_s$ and the capacitance $C_c$ between the contacts on the $K_3C_{60}$ film (See Supplementary Section S6.2).

To compare these ultrafast measurements to the DC response, the complex impedance $\hat{Z}$ was extracted from the complex valued $\overline{\Theta}(\omega, T) = \frac{2*Z_w + \hat{Z}(\omega,20K)}{2*Z_w + \hat{Z}(\omega,T)}$. In this expression, $Z_w$ denotes the wave impedance of the coplanar waveguide (see Fig. S12). The real part of $\hat{Z}(\omega, T)$ is displayed in Fig. 2c (blue dots), together with the two-point resistance measurement at DC (blue dashed line), yielding a qualitatively similar scaling.

Since these films are polycrystalline (see AFM image in Fig. S14), the equilibrium transition from the metallic state at T > $T_c$ to a purely inductive superconductor (T < 0.6 $T_c$) is broadened. This is understood by considering a granular superconductor [17-22, 33-36] composed of a set of grains undergoing a sharp superconducting transition at $T_c$, coupled to one another by weak links (Fig. 2c). Unlike the superconducting grains, these weak links exhibit finite resistivity down to appreciably lower temperature than $T_c$ due to thermally-activated phase slips, causing a gradual reduction in the overall resistance (see Supplementary Section S6.3).

The *nonlinear* electrical response of the sample, measured with the same ultrafast device of Fig 2a, confirms the picture of a granular superconductor qualitatively highlighted above. In these measurements, the peak current density of the probe pulses was controlled by changing the bias voltage of the leftmost launching switches. Fig. 3a shows representative $V_1(t)$ and $V_2(t)$ measurements at T = 8 K (<< $T_c$).

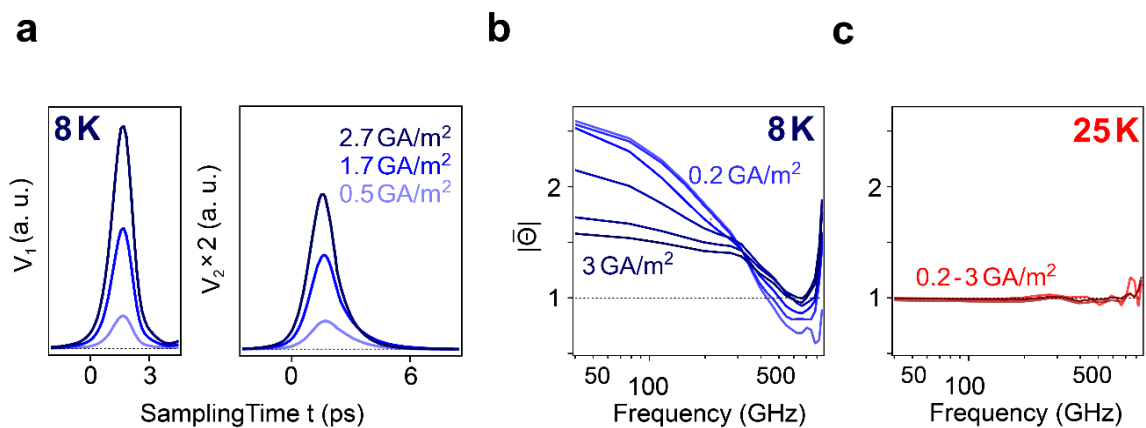

*Figure 3| Nonlinear ultrafast transport response of the equilibrium superconducting state.  **a,** Incoming and transmitted electrical pulse with different peak current density. The peak current density of the transmitted pulse is labelled. **b-c,** Current- and frequency-dependent normalized transmittance $|\bar{\theta}|$ at 8 K (**b**) and 25 K (**c**). The peak current density of the transmitted pulse is labelled. The black dashed lines indicate unity.*

The nonlinear transmittances of the superconducting state are shown in Fig. 3b, contrasted with measurements in the metallic state at 25 K (Fig. 3c). As expected, the transmittance of the equilibrium superconducting state reduces with increasing peak current density, whilst the equilibrium metallic state at 25 K exhibits no dependence on the current density.

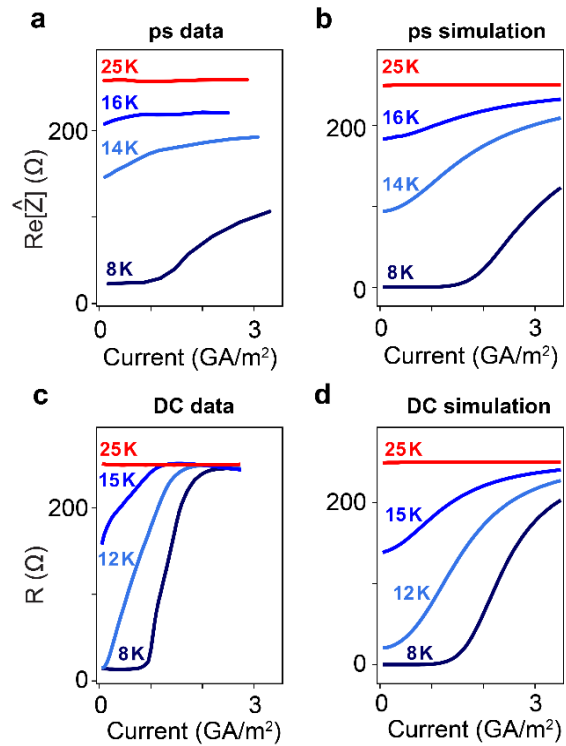

*Figure 4 | Critical current of the equilibrium superconducting state in picosecond and DC time scale. a-b, Ultrafast transport measurement (a) and simulation (b) of Re[$\hat{Z}$] at 50 GHz versus peak current density at 8, 14, 16 and 25 K. c-d, DC transport measurement (c) and simulation (d) of resistance versus current density at 8, 12, 15 and 25 K.*

These measurements are summarized by plotting the real part of the impedance at 50 GHz (Re[$\hat{Z}$]) as a function of the peak current density (Fig. 4a). At 8 K (dark blue curve), Re[$\hat{Z}$] remains low up to a critical peak current density $J_c \sim 1$ GA/m$^2$, above which the resistance starts to increase. For comparison, in Fig. 4c we report DC critical current measurements on the same device. The 8 K data of Fig 4a and 4c show a qualitatively similar response, with a similar critical current density value but a different rate of increase of the resistance above $J_c$. A qualitative similarity between picosecond and DC measurements was also observed for intermediate temperatures (0.6 $T_c$ < T < $T_c$), for which a finite resistance was detected at small currents.

All of these observations are explained by considering that the onset of resistive response is assigned to a combination of thermally generated and current induced phase slips at the weak links between superconducting grains. At the lowest temperatures, where the superconductor is well-formed and the DC resistivity vanishes (T = 8 K), these phase slips are infrequent and linear charge transport is dominated by inter-grain tunneling. When, at these lowest temperatures, the current density reaches $J_c$, phase slips are induced electrically [19-23], leading to a finite resistance that grows with bias current. The resistance growth is less steep for picosecond current pulses than for DC current bias, because the phase slips are induced only over the duration of the electrical pulse, and are less efficiently generated for short pulse durations (see Supplementary Section S5.1).

At intermediate temperatures where the superconducting grains are weakly coupled, phase slips are activated thermally already in absence of a strong current, causing a finite small-signal resistance [19-22]. With increasing currents, the resistance is increased by induced phase slips. This description is also verified by the scaling behavior $\Delta V \sim (I - I_c)^\alpha$ in DC transport measurements, where $\alpha$ changes from 1 to 3 when the temperature is reduced from 16 to 8 K (see Fig. S17) [22, 23, 37-40]. To compare DC and picosecond transport measurements, we simulated the measured data reported in Fig. 4a and 4c for both picosecond pulses and DC currents at different temperatures. The simulation was conducted by considering the equation of motion for Josephson junctions, similar to what was described in Ref. [17, 18]. The results are shown in Fig. 4b (picosecond simulation) and Fig. 4d (DC simulation), which reproduce the key features of the experiments (See Supplementary Section S6.3).

We next turn to the non-equilibrium photo-induced state, in search of transient signatures of superconductivity similar to those reported in Fig. 3 and Fig. 4. The optical-pump-electrical-probe delay $\tau$ (different from the sampling time $t$ of Fig. 2 and 3), was scanned by using a mechanical delay line for the pump beam. The data was obtained when the $K_3C_{60}$ film was held at T = 25 K and was excited with mid-IR ($\lambda$ = 7 μm) pulses at fluences up to 4 mJ/cm$^2$.

Similar to what was observed when cooling the sample below $T_c$, photo-excitation caused the transmitted pulse $V_2(t)$ to increase in amplitude, indicating a reduction in resistance. The normalized transmittances at two pump fluences 0.2 and 4 mJ/cm$^2$ are presented in Fig. 5b (more fluence-dependent data can be found in Supplementary Section 5.2). The key result is that the transmission of the photo-excited $K_3C_{60}$ film saturates to the level of the granular 16 K equilibrium response (T = 0.8 $T_c$), displayed for comparison in the plot (dashed lines). The calculated Re[$\hat{Z}$] at different pump fluences is shown in Fig. 5c, along with the equilibrium superconducting transition curve for comparison.

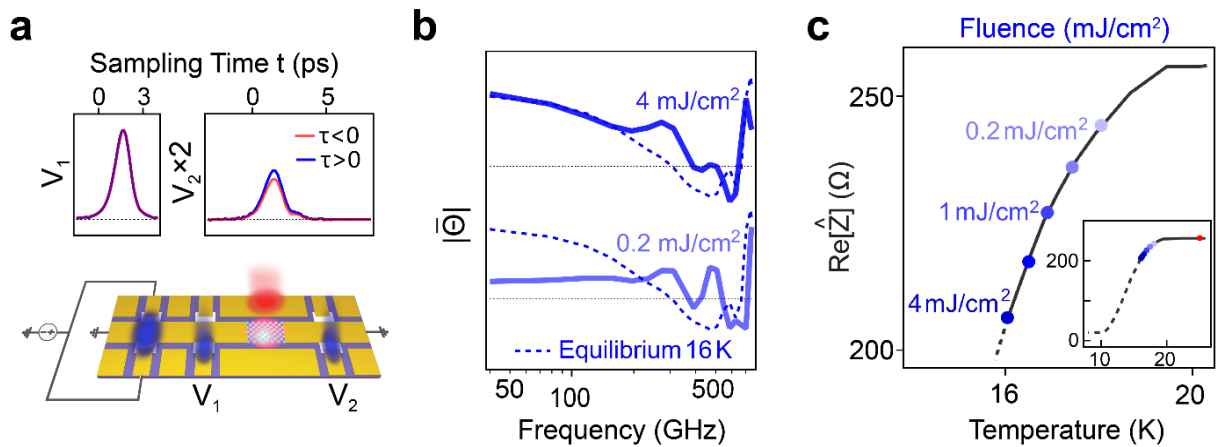

**Figure 5|Reduction of the impedance in the photo-induced state. a**, Lower: Schematic of the ultrafast transport measurement of the photo-induced state in $K_3C_{60}$. The red beam on the $K_3C_{60}$ thin film indicates the mid-IR excitation. Upper: Incoming pulse $V_1(t)$ and transmitted pulse $V_2(t)$ at $\tau < 0$ and $\tau > 0$. $\tau$ is the time delay between electronic pulse and mid-IR pulse, defined as $\tau = t_{e-pulse} - t_{mid-IR}$ where $t_{e-pulse}$ and $t_{mid-IR}$ are the arriving times of the electronic pulse and the mid-IR excitation, respectively. **b**, Normalized transmittances for pump fluences 0.2 and 4 mJ/cm$^2$ at 25 K. 4 mJ/cm$^2$ data is offset for clarity. The transmittance at equilibrium 16 K is plotted as blue dashed lines for comparison. The two black dashed lines indicate unity. **c**, Dependence of Re[$\hat{Z}$] on pump fluence in the photo-induced state (upper axis). Dots from right to left (lighter to darker blue) are measurements at 0.2, 0.5, 1, 2 and 4 mJ/cm$^2$. Re[$\hat{Z}$] versus temperature (lower axis) in the equilibrium state is

*shown by a black line for comparison. Here pump-probe measurements are taken with mid-IR pump at T = 25 K and at τ = 5 ps. Inset: Reduction of Re[$\hat{Z}$] of the photo-induced state compared with the equilibrium superconducting transition. The red dot indicates the position of the unpumped state at 25 K.*

The same conclusion was drawn when probing the transient *nonlinear* transport properties. Figure 6a displays the normalized transmittance of the equilibrium state (left) and the photo-induced state (right) at different bias currents. Just as in the transient linear impedance, the nonlinear transport properties of the photo-induced state showed the same behavior as that observed in equilibrium at 16 K.

The observation of a photo-induced state with nonlinear I-V characteristics is uniquely indicative of non-equilibrium superconductivity. In fact, when the same film was excited with visible light pulses (515 nm), for which no photo-induced superconductivity is expected [12], this nonlinearity disappeared. This comparison is summarized in Fig. 6b, where the corresponding changes of Re[$\hat{Z}$] are plotted as a function of peak current density. In this panel, the nonlinear electrical transport of the photo-induced superconducting state is compared with the equilibrium granular superconductor at T = 16 K, and contrasted with the response of the equilibrium metal and of that excited in the visible, for which no nonlinearity was observed.

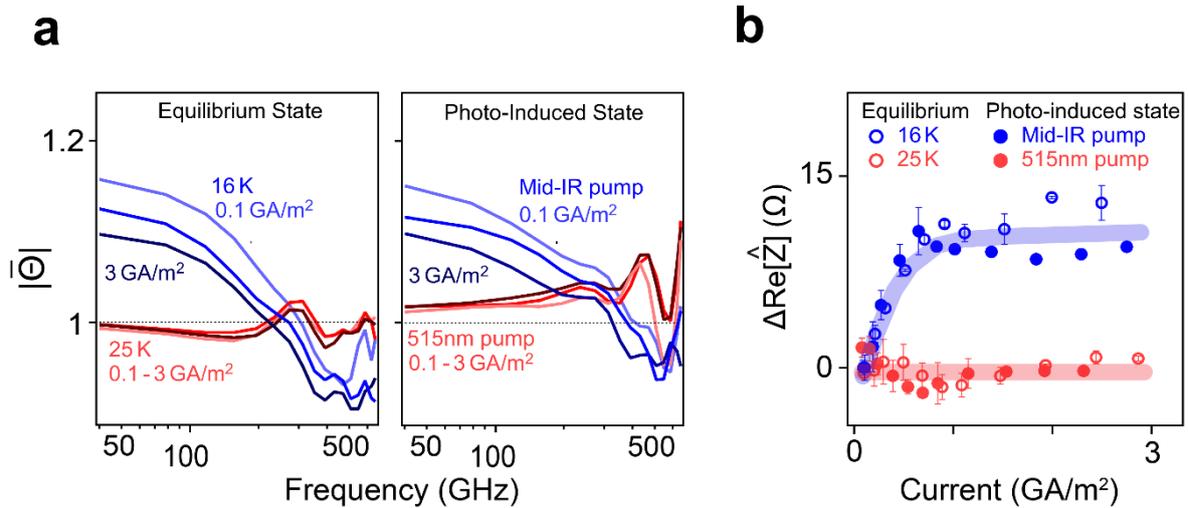

*Figure 6 | Critical current of the photo-induced state. **a**, Left: Current dependence of the transmittance in equilibrium at 16 K (blue) and at 25 K (red). Right: Current dependence of the transmittance of the photo-induced state with mid-IR excitation (blue) and 515 nm excitation (red). **b**, Change of the real part of the impedance Re[$\hat{Z}$] with increasing current in equilibrium at 16 K (blue circle), at 25 K (red circle) and for the photo-induced state with mid-IR excitation (blue dots) and 515 nm excitation (red dots).*

In seeking to interpret the nature of the photo-induced state, we propose that the effect of the mid-infrared pump is to turn the grains into a transient superconducting state, whilst the weak links connecting these regions are left in an excited state. This excitation at the weak links may be a result of energy deposited in the films, or even of a fast quench which drives the order parameter phase to different values in neighboring grains. Hence, the observation that even at the highest fluences we could not induce properties observed in equilibrium below T=16 K (0.8 $T_c$) may be a result of the polycrystalline nature of the sample and of weak links that *heat* upon photo-excitation. It is not clear that the pressed powders used in the previously reported optical studies would exhibit the same type of transient granular superconductivity documented here. Systematic work is needed for different film configurations and thicknesses, as well as further optimization of the growth process.

This interpretation appears more likely than one in which the finite residual resistance of the photo-induced state is attributed to a fraction of the sample becoming superconducting after excitation, leaving another fraction metallic. This is predicated on the change in resistance versus temperature at different bias currents (See Fig. S16), which does not exhibit double transition features [41-44] and indicates the stoichiometric homogeneity of the thin-film sample. Secondly, the fact that the sample is pumped by an optical beam with a size similar to the sample size (FWHM of pump beam ~ 21 μm and sample dimensions 20 x 20 μm²), makes an inhomogeneous photo-induced state less convincing at this stage.

In summary, we have developed a new experimental platform to probe ultrafast nonlinear transport in doped fullerite thin films. We report unique signatures of photo-induced superconductivity as nonlinear I-V characteristics in the transient state. In these devices, which are based on polycrystalline films, the photo-induced state never acquires the transport properties of the homogeneous superconductor and rather saturates into a non-equilibrium with both *linear* and *non-linear* transport properties similar to those observed in equilibrium for $T \sim 0.8\ T_c$. This is possibly a result of the geometry of the samples, although we cannot exclude that under all circumstances, fast quenches may leave a photo-induced state with a high density of vortex excitations and other topological defects. Future work could be extended to different superconductors, whether by using new growth protocols or by exfoliating layered materials such as Bi-based cuprates and van der Waals based superconductors. These studies will also make accessible a wider variety of non-equilibrium phenomena in quantum materials, including sliding charge and spin density waves as well as domain wall propagation. Finally, the work reported here represents an important step towards integration of quantum materials into ultrafast electrical devices, which may lead to high-bit rate applications based on the functionalities of driven quantum materials.

**REFERENCES**

1. Radu, I., et al., *Transient ferromagnetic-like state mediating ultrafast reversal of antiferromagnetically coupled spins.* Nature, 2011. **472**(7342): p. 205-208.
2. Wang, Y.H., et al., *Observation of Floquet-Bloch States on the Surface of a Topological Insulator.* Science, 2013. **342**(6157): p. 453-457.
3. Nova, T.F., et al., *Metastable ferroelectricity in optically strained $SrTiO_3$.* Science, 2019. **364**(6445): p. 1075-1079.
4. Li, X., et al., *Terahertz field induced ferroelectricity in quantum paraelectric $SrTiO_3$.* Science, 2019. **364**(6445): p. 1079-1082.
5. Disa, A.S., et al., *Polarizing an antiferromagnet by optical engineering of the crystal field.* Nature Physics, 2020. **16**(9): p. 937-941.
6. McIver, J.W., et al., *Light-induced anomalous Hall effect in graphene.* Nature Physics, 2020. **16**(1): p. 38-41.
7. Wang, X., et al., *Light-induced ferromagnetism in moiré superlattices.* Nature, 2022. **604**(7906): p. 468-473.
8. Fausti, D., et al., *Light-Induced Superconductivity in a Stripe-Ordered Cuprate.* Science, 2011. **331**(6014): p. 189-191.
9. Hu, W., et al., *Optically enhanced coherent transport in $YBa_2Cu_3O_{6.5}$ by ultrafast redistribution of interlayer coupling.* Nature Materials, 2014. **13**(7): p. 705-711.
10. Cremin, K.A., et al., *Photoenhanced metastable c-axis electrodynamics in stripe-ordered cuprate $La_{1.885}Ba_{0.115}CuO_4$.* Proceedings of the National Academy of Sciences, 2019. **116**(40): p. 19875-19879.
11. Isoyama, K., et al., *Light-induced enhancement of superconductivity in iron-based superconductor $FeSe_{0.5}Te_{0.5}$.* Communications Physics, 2021. **4**(1): p. 160.
12. Mitrano, M., et al., *Possible light-induced superconductivity in $K_3C_{60}$ at high temperature.* Nature, 2016. **530**(7591): p. 461-464.
13. Cantaluppi, A., et al., *Pressure tuning of light-induced superconductivity in $K_3C_{60}$.* Nature Physics, 2018. **14**(8): p. 837-841.
14. Buzzi, M., et al., *Photomolecular High-Temperature Superconductivity.* Physical Review X, 2020. **10**(3): p. 031028.
15. Budden, M., et al., *Evidence for metastable photo-induced superconductivity in $K_3C_{60}$.* Nature Physics, 2021. **17**(5): p. 611-618.
16. Buzzi, M., et al., *Phase Diagram for Light-Induced Superconductivity in $k$-$(ET)_2$-X.* Physical Review Letters, 2021. **127**(19): p. 197002.
17. Ambegaokar, V. and B.I. Halperin, *Voltage Due to Thermal Noise in the dc Josephson Effect.* Physical Review Letters, 1969. **22**(25): p. 1364-1366.
18. Falco, C.M., et al., *Effect of thermal noise on current-voltage characteristics of Josephson junctions.* Physical Review B, 1974. **10**(5): p. 1865-1873.
19. Halperin, B.I. and D.R. Nelson, *Resistive transition in superconducting films.* Journal of Low Temperature Physics, 1979. **36**(5): p. 599-616.
20. Resnick, D.J., et al., *Kosterlitz-Thouless Transition in Proximity-Coupled Superconducting Arrays.* Physical Review Letters, 1981. **47**(21): p. 1542-1545.
21. Lobb, C.J., D.W. Abraham, and M. Tinkham, *Theoretical interpretation of resistive transition data from arrays of superconducting weak links.* Physical Review B, 1983. **27**(1): p. 150-157.
22. England, P., et al., *Granular superconductivity in $R_1Ba_2Cu_3O_{7-\delta}$ thin films.* Physical Review B, 1988. **38**(10): p. 7125-7128.
23. Tinkham, M., *Introduction to Superconductivity*. 2004: Dover Publications.
24. Hebard, A.F., et al., *Superconductivity at 18 K in potassium-doped $C_{60}$.* Nature, 1991. **350**(6319): p. 600-601.
25. Xiang, X.-D., et al., *Synthesis and Electronic Transport of Single Crystal $K_3C_{60}$.* Science, 1992. **256**(5060): p. 1190-1191.

26. Takabayashi, Y., et al., *The Disorder-Free Non-BCS Superconductor $Cs_3C_{60}$ Emerges from an Antiferromagnetic Insulator Parent State.* Science, 2009. **323**(5921): p. 1585-1590.
27. Ganin, A.Y., et al., *Polymorphism control of superconductivity and magnetism in $Cs_3C_{60}$ close to the Mott transition.* Nature, 2010. **466**(7303): p. 221-225.
28. Gunnarsson, O., *Alkali-Doped Fullerides*. Alkali-Doped Fullerides.
29. Takabayashi, Y. and K. Prassides, *Unconventional high $T_c$ superconductivity in fullerides.* Philosophical Transactions of the Royal Society A: Mathematical, Physical and Engineering Sciences, 2016. **374**(2076): p. 20150320.
30. Auston, D.H., *Picosecond optoelectronic switching and gating in silicon.* Applied Physics Letters, 1975. **26**(3): p. 101-103.
31. Prechtel, L., et al., *Spatially resolved ultrafast transport current in GaAs photoswitches.* Applied Physics Letters, 2010. **96**(26): p. 261110.
32. Gallagher, P., et al., *Quantum-critical conductivity of the Dirac fluid in graphene.* Science, 2019. **364**(6436): p. 158-162.
33. Falco, C.M., W.H. Parker, and S.E. Trullinger, *Observation of a Phase-Modulated Quasiparticle Current in Superconducting Weak Links.* Physical Review Letters, 1973. **31**(15): p. 933-936.
34. Beasley, M.R., J.E. Mooij, and T.P. Orlando, *Possibility of Vortex-Antivortex Pair Dissociation in Two-Dimensional Superconductors.* Physical Review Letters, 1979. **42**(17): p. 1165-1168.
35. Baenitz, M., et al., *Inter- and intragrain AC response of the granular superconductors $K_3C_{60}$ and $Rb_3C_{60}$.* Physica C: Superconductivity, 1994. **228**(1): p. 181-189.
36. Buntar, V., *Investigation of inter- and intragrain currents in $K_3C_{60}$ single crystals.* Physica C: Superconductivity, 1998. **309**(1): p. 98-104.
37. Berezinskiĭ, V.L., *Destruction of Long-range Order in One-dimensional and Two-dimensional Systems having a Continuous Symmetry Group I. Classical Systems.* Soviet Journal of Experimental and Theoretical Physics, 1971. **32**: p. 493.
38. Berezinskiĭ, V.L., *Destruction of Long-range Order in One-dimensional and Two-dimensional Systems Possessing a Continuous Symmetry Group. II. Quantum Systems.* Soviet Journal of Experimental and Theoretical Physics, 1972. **34**: p. 610.
39. Kosterlitz, J.M. and D.J. Thouless, *Ordering, metastability and phase transitions in two-dimensional systems.* Journal of Physics C: Solid State Physics, 1973. **6**(7): p. 1181.
40. Abraham, D.W., et al., *Resistive transition in two-dimensional arrays of superconducting weak links.* Physical Review B, 1982. **26**(9): p. 5268-5271.
41. Gerber, A., et al., *Nonmonotonic resistivity transitions in granular superconducting ceramics.* Physical Review B, 1991. **43**(16): p. 12935-12942.
42. Jardim, R.F., et al., *Granular behavior in polycrystalline $Sm_{2-x}Ce_xCuO_{4-y}$ compounds.* Physical Review B, 1994. **50**(14): p. 10080-10087.
43. Gerber, A., et al., *Double-peak superconducting transition in granular L-M-Cu-O(L=Pr,Nd,Sm,Eu,D; M=Ce,Th) superconductors.* Physical Review Letters, 1990. **65**(25): p. 3201-3204.
44. Early, E.A., et al., *Double resistive superconducting transition in $Sm_{2-x}Ce_xCuO_{4-y}$.* Physical Review B, 1993. **47**(1): p. 433-441.

# Supplementary information for

# Nonlinear transport in a photo-induced superconductor

E. Wang[1], J. D. Adelinia[1], M. Chavez-Cervantes[1], T. Matsuyama[1], M. Fechner[1], M. Buzzi[1], G. Meier[1], A. Cavalleri[1,2]

[1] *Max Planck Institute for the Structure and Dynamics of Matter, Hamburg, Germany*

[2] *Department of Physics, Clarendon Laboratory, University of Oxford, United Kingdom*

# Contents

# S1. Experimental setup

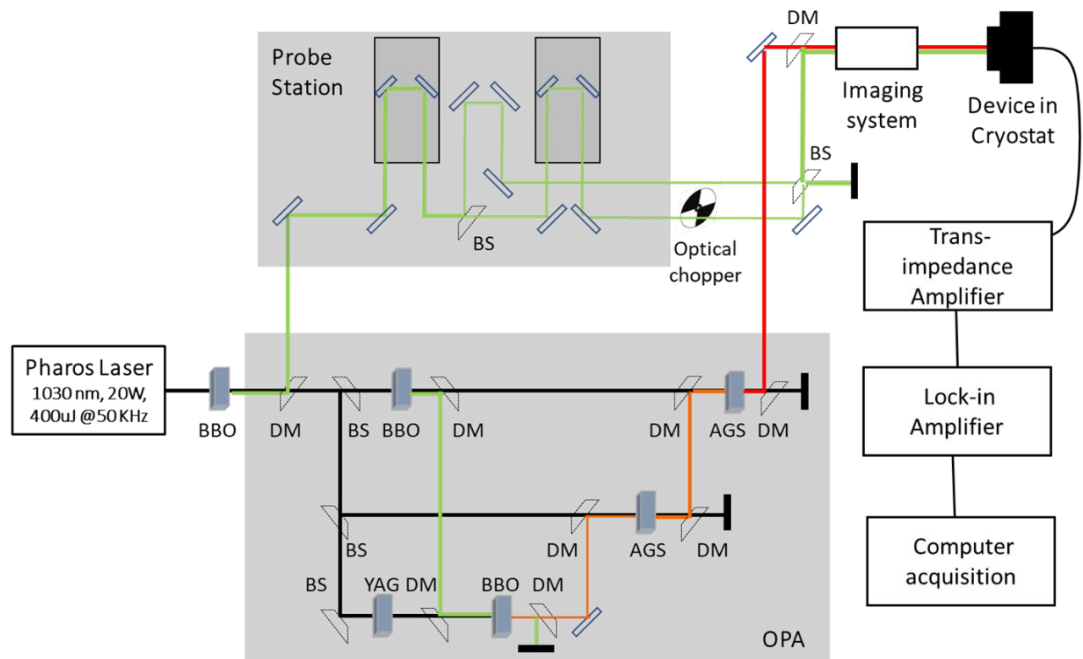

*Fig. S 1| Schematics of the experimental setup. Abbreviations of objects: BBO-Beta Barium Borate crystal; DM-Dichroic Mirror; BS-Beam Splitter; YAG-Yttrium Aluminum Garnet crystal; AGS-Silver Thiogallate crystal; OPA-Optical Parametric Amplifier. The wavelength of optical pluses is represented by color: 515 nm (green), 1030 nm (black), 1200 nm (orange) and 7 µm (red).*

The optical part of the experimental setup is based on a Pharos Laser, which provides optical pulses with pulse energy 400 µJ and duration 250 fs, with a wavelength centered at 1030 nm and a repetition rate of 50 kHz (Fig. S 1). The output pulses were first sent through a 0.1 mm thick Beta Barium Borate (BBO) crystal to generate visible pulses with wavelength 515 nm. The 515 nm pulses and the leftover 1030 nm pulses were separated using a dichroic mirror (DM). The 515 nm pulses were used to excite photo-conductive switches on the sample device. The leftover 1030 nm pulses were used to power a optical parameter amplifier (OPA) utilizing silver thiogallate crystals based on the design of reference [1]. The mid-IR pulses (7 µm) from the OPA and the 515 nm pulses were directed into a Janis ST-500 optical cryostat through an imaging system based on a pellicle beam splitter and a reflective objective to focus them onto the device. The wavelength of the mid-IR pulses was characterized by Fourier Transform InfraRed measurements (Fig. S 2) and the beam diameter in the focal plane was measured using a typical knife-edge method (Fig. S 3).

During all measurements, the electrical signals from the photo-conductive switches were amplified by an in-house custom-built transimpedance amplifier via electrical feedthroughs from the cryostat and then measured using a lock-in amplifier. The 515 nm light used to launch the electrical pulses was chopped at approximately 1kHz, as indicated in Fig. S 1, and this frequency was used as the reference frequency of the lock-in amplifier. Optical delay lines were used to scan the mutual time delay between the launching and the sampling 515 nm pulses, and to adjust the delay of the 7 μm pulses relative to both 515 nm pulses. The data was recorded using custom LabVIEW software.

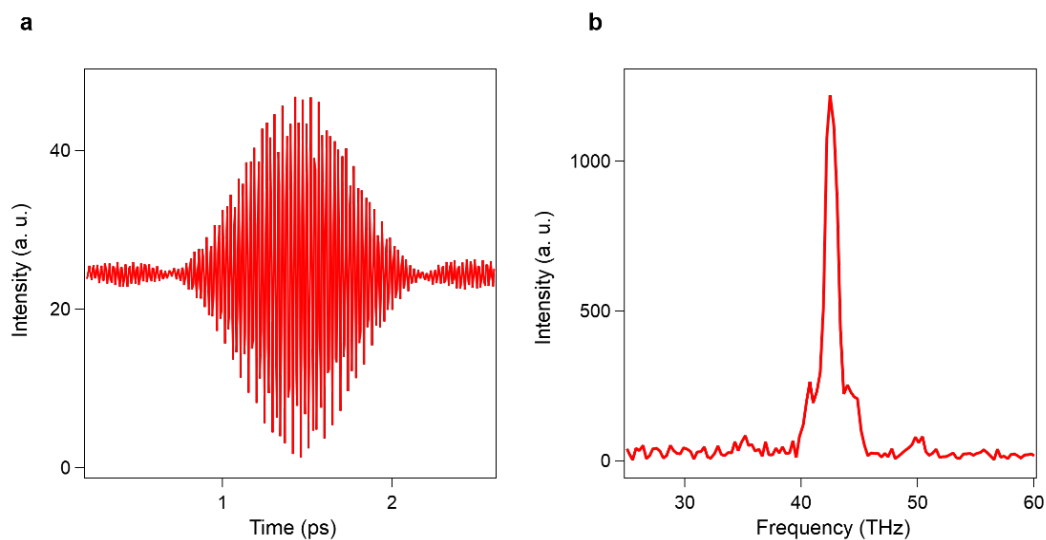

Fig. S 2| *Fourier Transform InfraRed (FT-IR) spectrum of the mid-IR. (a) Time trace of the interference signal from the Michelson interferometer. (b) Fast Fourier transform of the time trace in (a).*

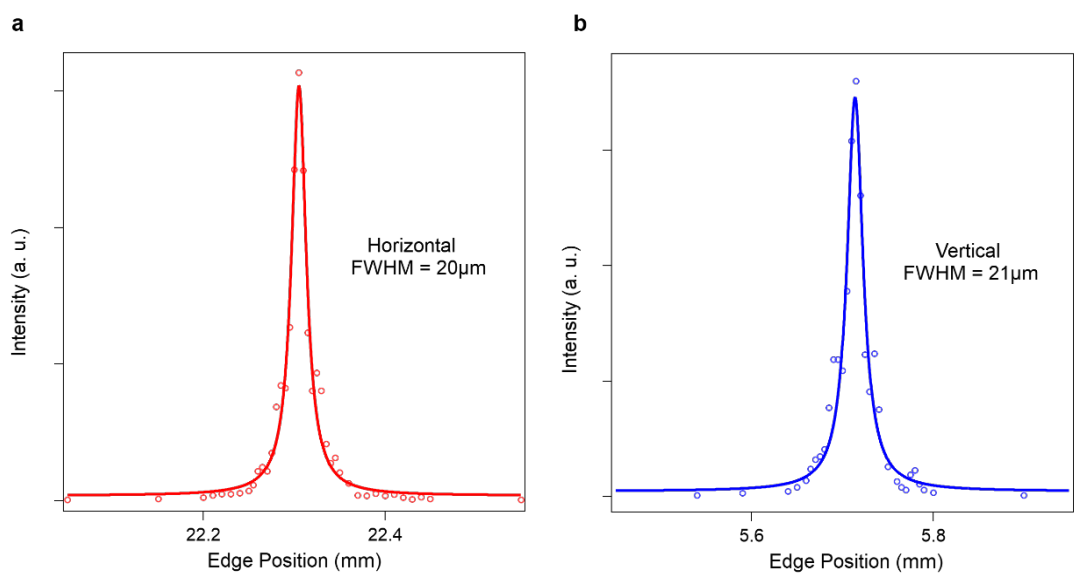

Fig. S 3| *Horizontal and vertical beam size of the mid-IR pulses at the focal plane of the imaging system, measured using the knife-edge method.*

# S2. Device fabrication process

Prior to $K_3C_{60}$ thin film growth, the photo-conductive switches and coplanar waveguide were fabricated on a sapphire substrate using laser lithography and electron-beam evaporation. This was done in two steps – first the switches, then the waveguide. For both steps, a bilayer photoresist mask was used for the lithography, for which MicroChem LOR-7B served as an undercut layer and micro resist ma-P 1205 served as the light-sensitive top layer. After developing the structure, a 200 nm layer of silicon was evaporated for the photo-conductive switches (Fig. S 4a). In the second step, 10 nm of titanium immediately followed by 280 nm of gold was evaporated to form the coplanar waveguide structure (Fig. S 4b).

Afterwards, a shadow mask with a 20 μm x 20 μm square hole in the center was aligned to the middle of the device under an optical microscope using a micro-manipulator (Fig. S 4c). The whole device was then transferred to a molecular beam epitaxy (MBE) chamber and was degassed at 300 °C for 12 hours prior to growth. The $C_{60}$ molecules were deposited on the device from an effusion cell source with the device kept at 200 °C and the $C_{60}$ source at 380 °C. The deposition rate was ~ 1 nm/min. The thickness of the $C_{60}$ thin film was ~100 nm. During growth, the temperature of the device was monitored using a pyrometer.

To improve the electrical contact between the $C_{60}$ thin film and the signal line of the waveguide, 10 nm Ti/350 nm Au were deposited additionally onto the contact area (Fig. S 4d).

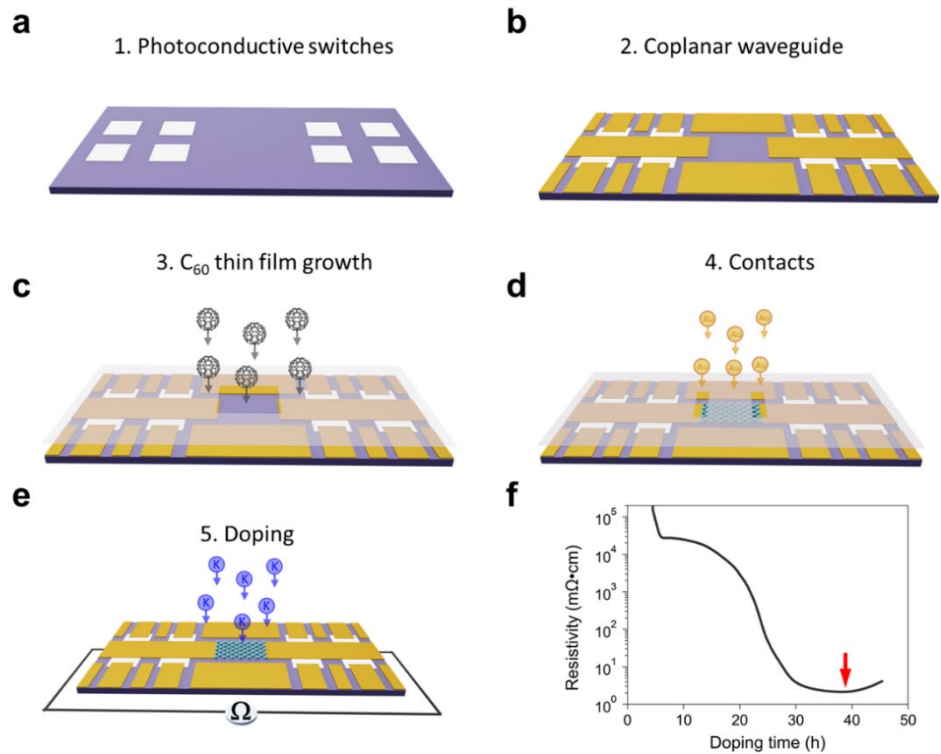

*Fig. S 4| Schematics of device fabrication steps (a) to (e). (f) Resistivity profile of the doping process.*

To obtain a $K_3C_{60}$ film, potassium was deposited on the $C_{60}$ thin film from an effusion cell source with the device kept at 200 °C and the potassium source at 100 °C. To precisely control the stoichiometry, the device was connected to an ohmmeter via electrical feedthroughs from the chamber and the resistance of film was monitored in-situ during the doping process (Fig. S 4e). The resistance decreases upon initial doping and subsequently reaches a minimum (indicated by the red arrow in Fig. S 4f), which corresponds to the formation of $K_3C_{60}$ as reported in previous studies [2, 3]. A superconducting transition with $T_c$ ~ 19 K as shown in the main text and below further confirms the correct stoichiometry. In order to achieve homogeneous doping, a dope-anneal cycle consisting of 1 hour of doping followed by a 6-hour anneal was used to allow enough time for potassium to diffuse inside the film.

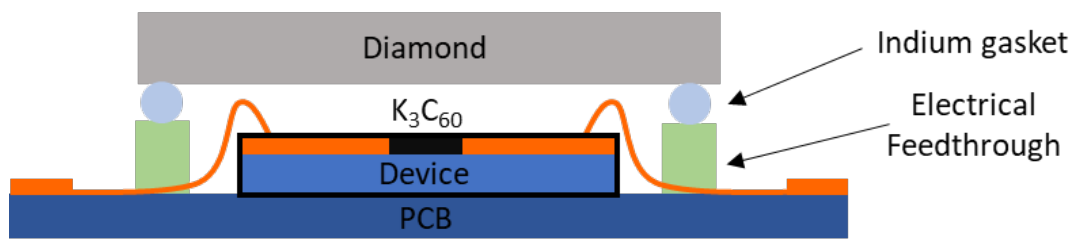

*Fig. S 5| Schematics of the device after sealing.*

After doping, the device was transferred to a glovebox using a high-vacuum suitcase to avoid oxidation. In the glovebox, the device was attached to a home-made printed circuit board (PCB), and sealed with a diamond window together with an indium gasket (Fig. S 5). The electrical pads on the device were connected to the PCB via electrical feedthroughs fabricated with low vapor pressure epoxy Torr Seal. After sealing, the device was transferred into the optical cryostat for measurements. During the measurements, the optical pulses were incident on the device through the top diamond window.

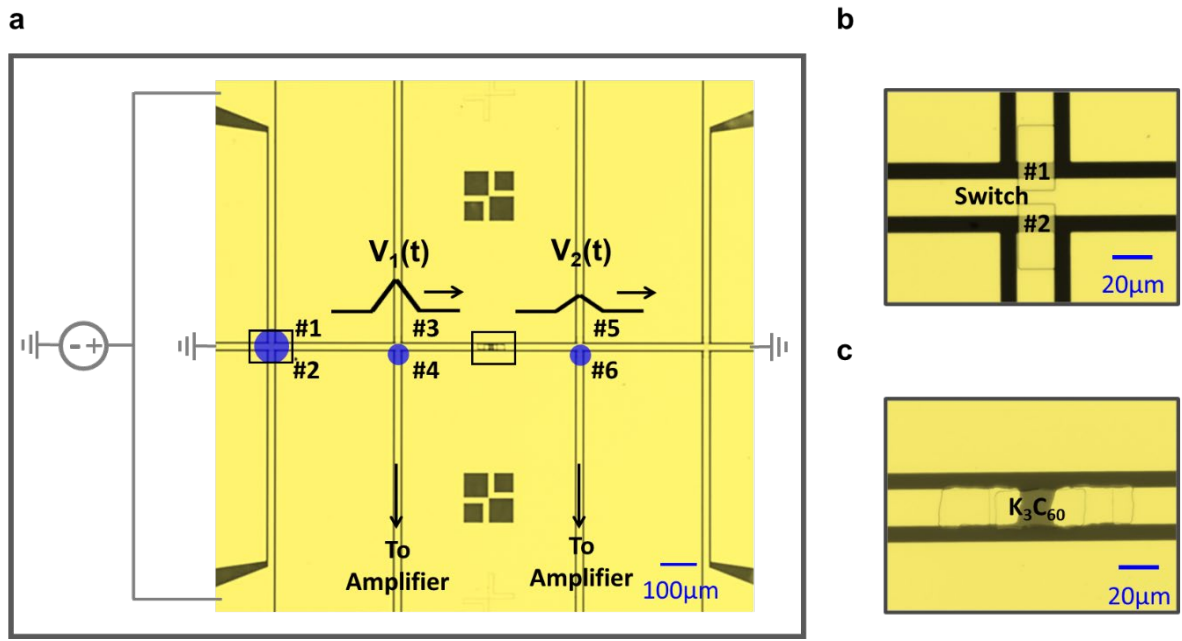

Fig. S 6| *Optical microscope image of the device. #1-#6 denote the three pairs of photo-conductive switches. $V_1(t)$ and $V_2(T)$ denote the reference and transmitted electrical pulse. (b, c) Zoom-in images of a pair of photo-conductive switches (b) and $K_3C_{60}$ thin film (c) as indicated by black boxes in (a).*

A microscope image of the device is shown in Fig. S 6. The three pairs of photo-conductive switches are labelled #1 to #6. The zoom-in images of switches and $K_3C_{60}$ thin film are shown in Fig. S 6b and c. During the measurements, switches #1 and #2 were biased simultaneously and were illuminated together with 515 nm pulses (with fluence ~ 4 mJ/cm²), to launch the electrical pulse $V_1(t)$. $V_1(t)$ was sampled at switch #4 with another 515 nm pulse (~4 mJ/cm²). After measuring $V_1(t)$, the transmitted electrical pulse $V_2(t)$ was sampled by illuminating switch #6 instead of switch #4. The corresponding electrical signals from both detection switches were collected using transimpedance and lock-in amplifiers.

## S3. Device and sample characterization

S3.1: Characterization of the launched electrical pulse

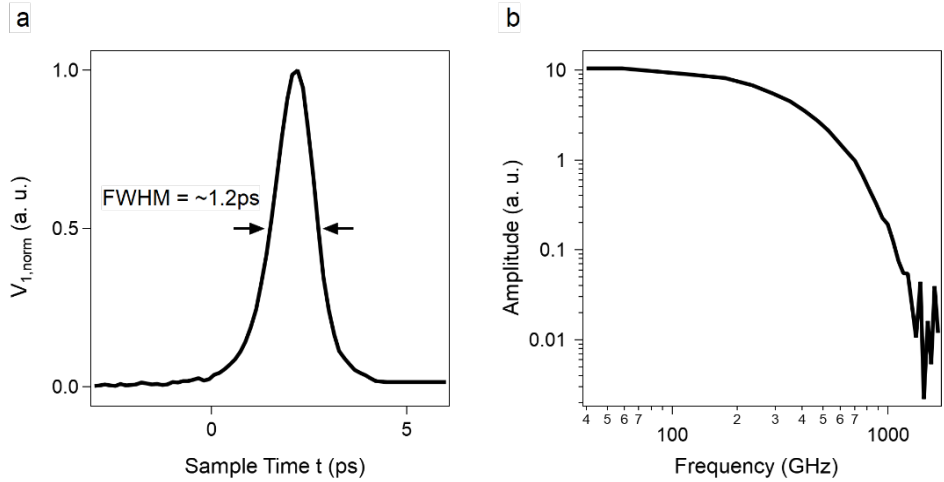

Fig. S 7|Temporal width and spectrum of the launched electrical pulse. (a) Time trace of the launched electrical pulse, which is the reference pulse $V_1(t)$ in experiment. The shown curve is normalized to unity. (b) Fast Fourier transform of the time trace in (a) to show the spectrum in frequency domain.

The time trace of the electrical pulse launched from photo-conductive switches #1 and #2, is shown in Fig. S 7a. The full width at half maximum (FWHM) is ~1.2 ps and the spectral weight of pulse is very weak above 1 THz (Fig. S 7b). The limiting factors for the pulse width are the lifetime of photo-carriers in the switches and the bandwidth of coplanar waveguide used in the experiment. These will be discussed in the following sections S3.2 and S3.5.

## S3.2: Lifetime of photo-carriers in the photo-conductive switches

The lifetime of the photo-carriers in the photo-conductive switches was determined by measuring the auto-correlation curves of the switch response functions (proportional to the time profile of the photo-carrier population). During the measurement, we illuminated switches #3 and #4 with two small-diameter 515 nm pulses which each only covers one switch. Meanwhile, one switch was biased with a voltage and the other one remained unbiased (Fig. S 8a). By changing the mutual delay between the two 515 nm pulses (denoted as sampling time t), a time profile was measured which was the auto-correlation of the switch response functions (Fig. S 8c). The measured time profiles are simulated by assuming the response function has a rising time the same as the pulse duration of the 515 nm pulse (~ 250 fs) and a decay time the same as the photo-carrier lifetime $\tau_e$ (Fig. S 8b). Fig. S 8c shows the comparison between measurements and different simulation results with various $\tau_e$ values. The photo-carrier lifetime $\tau_e$ is estimated to be between 250 – 300 fs. This result was observed to be independent of temperature between 8 and 50 K.

The photo-carrier lifetime in our experiment is shorter than that observed in previous studies [4]. This is possibly caused by the annealing process before $C_{60}$ deposition, which could cause the migration of titanium or gold atoms into the silicon patches and increase the scattering rate.

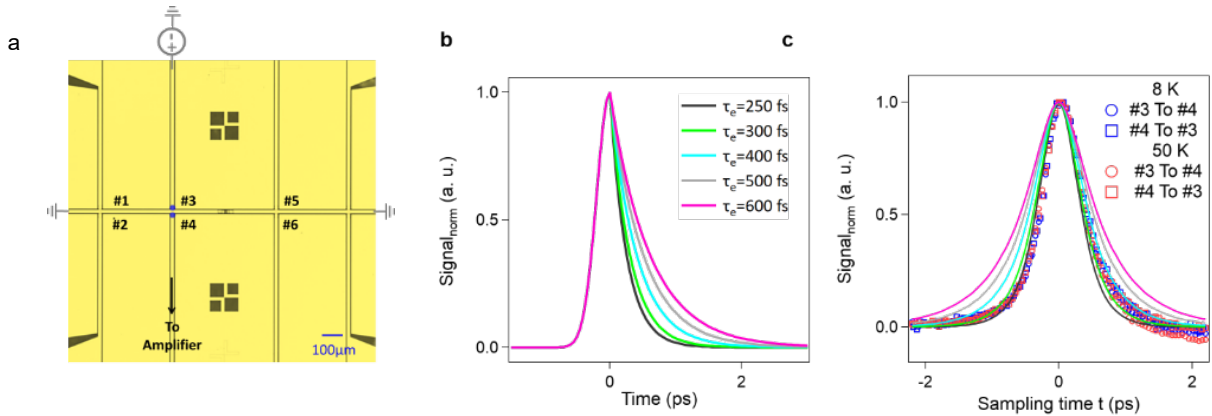

Fig. S 8| Lifetime of the photo-carriers in the photo-conductive switches. (a) Setup of measurement. (b) Response function of the photo-conductive switch, with a rising time of ~ 250 fs, for various photo-carrier lifetimes $\tau_e$. (c) Comparison of the auto-correlation of the response function and the measurements by exciting switches #3 and #4. The geometry in (a) corresponding to measurement #3 to #4, in which switch #3 is biased and #4 is connected to transimpedance amplifier. In measurement #4 to #3, this is reversed.

## S3.3: Calibration of photo-conductive switches

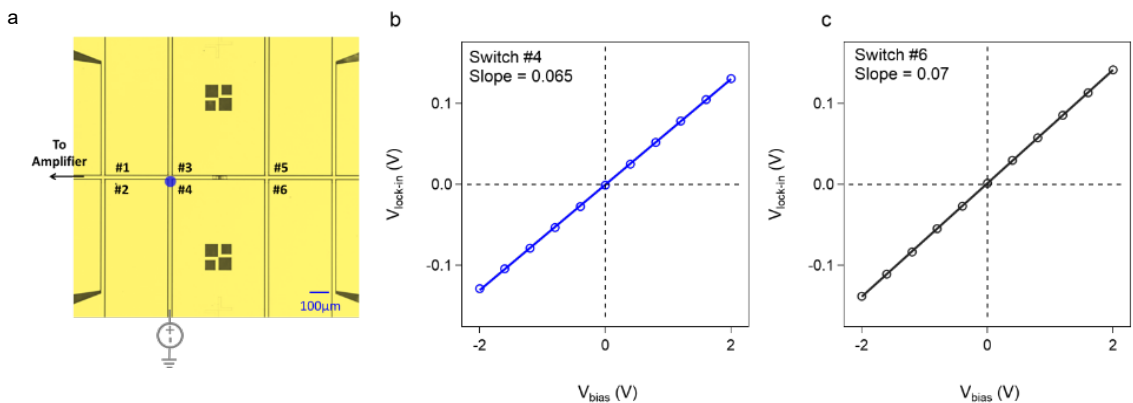

Fig. S 9| Calibration of the photo-conductive switches. (a) Setup for calibration process for switch #4. The switch being calibrated was biased with a voltage and illuminated with 515 nm pulses. (b, c) The signal measured with the lock-in amplifier versus the bias voltage. The slope was calculated for each switch and used as the normalization factor individually.

During the experiments, it was difficult to ensure that the 515 nm pulses used to illuminate all switches were identical, which means that during the measurements each switch had a slightly different sensitivity. To calibrate this difference, before each measurement, the switch that was used to probe the electrical pulse was calibrated independently. The switch being calibrated was biased with a voltage and illuminated with 515 nm pulses. The generated electrical signal was collected at one end of the signal line (the other end was left open) through the combination of a transimpedance amplifier and a lock-in amplifier (Fig. S 9a). The dependence of the signal on the bias voltage is shown in Fig. S 9(b, c) for the two switches which were used to sample the reference pulse $V_1(t)$ and the transmitted pulse $V_2(t)$, respectively. The sensitivity of the switch is directly represented by the slope of the curve which was used as a normalization factor to normalize the data in our experiment. For each measurement in the experiment, the sensitivity of the switch which was used to sample the electrical pulse was calibrated beforehand.

S3.4: Launching of an ultrashort quasi-TEM electrical pulse

In the experiment, it is important that the launched ultrashort electrical pulse has a single quasi-TEM mode, as different excitation modes have different impedance, which would make the analysis difficult. In our case with the coplanar waveguide, switches #1 and #2 (Fig. S 6) were biased simultaneously and symmetrically using a voltage source, and were illuminated uniformly. In this way, the electrical pulse was launched in a symmetrical way as even mode (meaning the electrical field is symmetrical about the signal line), which is the quasi-TEM mode in the coplanar waveguide. To confirm, we measured the electrical pulse with two switches #3 and #4, which are on the upper and lower side at the same position of the signal line. The measurements yielded almost identical signals as shown in Fig. S 10, indicating that we launch an even mode into the waveguide with this technique.

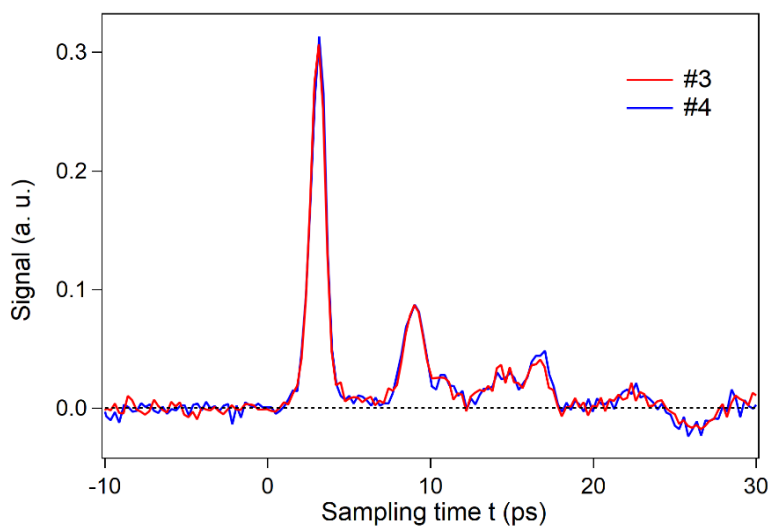

*Fig. S 10| Characterization of the launched electrical pulse. The launched electrical pulse was sampled using switches #3 and #4. The signals are proportional to the electrical field on the two sides of signal line.*

S3.5: Characterization of the coplanar waveguide: Damping and impedance

Other than the photo-carrier lifetime, another limiting factor to the temporal width of generated electrical pulse is the damping caused by the coplanar waveguide used in the experiment. The frequency-dependent transmittance of the quasi-TEM pulse for a 200 μm long waveguide, extracted from a finite-element time-domain simulation in CST studio suite, is shown in Fig. S 11. The frequency component above 1 THz is strongly damped, with the -6 dB point at approximately 1 THz. This is consistent with the electrical pulse width of ~ 1 ps observed in the experiment.

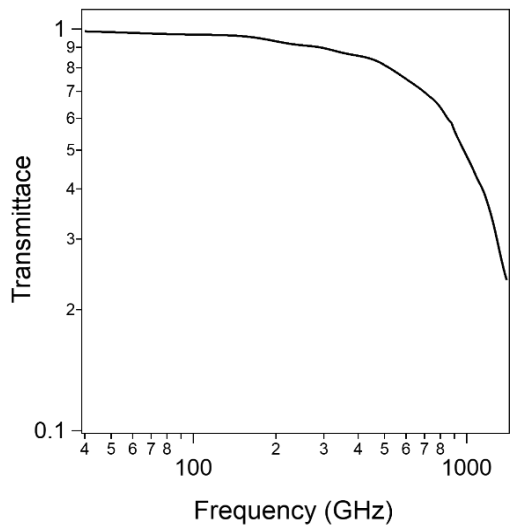

*Fig. S 11| Frequency domain finite element simulation of the transmittance of the 200 μm long terahertz coplanar waveguide used in the experiment.*

The frequency-dependent real and imaginary parts of the impedance of the coplanar waveguide $Z_w$ for the quasi-TEM pulse, calculated using CST studio suite, are shown in Fig. S 12. Within the frequency range (40-500 GHz) relevant for present experiments, both Re[Z] and Im[Z] exhibit a weak frequency dependence. Re[Z] is around 59 Ω and is much larger than Im[Z] (|Im[Z]| < 0.2 Ω). Therefore, in the calculation for the impedance of the $K_3C_{60}$ film, we use a purely real impedance for the waveguide, $Z_w$=59 Ω.

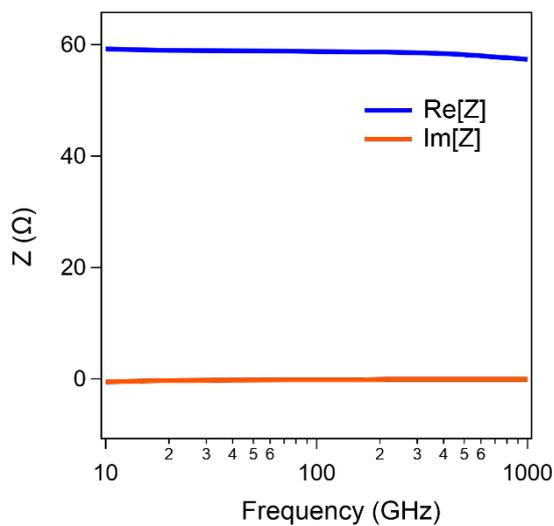

*Fig. S 12| Simulated frequency dependence of the real and imaginary parts of the impedance of the terahertz coplanar waveguide used in the experiment.*

## S3.6: Calibration of peak current value of ultrashort electrical pulse

To calibrate the peak current value of the ultrashort electrical pulse, the net charge of the electrical pulse was collected at one end of the signal line, which was measured using a combination of a transimpedance amplifier and a lock-in amplifier, as shown in Fig. S 13. The peak current value is calculated with the relation:

$$I_{\text{peak}} = \frac{1}{2} \cdot \frac{2\sqrt{2}}{1.273} \cdot \frac{V_{\text{Lock-in}}}{A_{\text{TIA}}} \cdot \frac{1}{f_{\text{rep}}} \cdot \frac{1}{\int pulse_{\text{norm}}(t)dt}$$

Where $I_{\text{peak}}$ is the current peak value, $A_{\text{TIA}} (= 2 \cdot 10^9 \frac{V}{A})$ is the amplification factor of transimpedance amplifier, $f_{\text{rep}} (= 50\ kHz)$ is the laser repetition rate, and $\int pulse_{\text{norm}}(t)dt$ ($\sim 1.5 \cdot 10^{-12} s$) is the temporal integral of the normalized electrical pulse. The factor $1/2$ comes from the fact that two equal electrical pulses propagate towards left and right direction. The factor $2\sqrt{2}/1.273$ describes the relation between the net charge amplitude and the root mean square (RMS) amplitude measured at the lock-in amplifier at the chopping frequency. The calculated peak current value of reference pulse ($V_1(t)$) versus bias voltage is shown in Fig. S 13b. In the main text, the data are plotted versus the peak current value of the transmitted pulse $V_2(t)$, which is $I_{\text{peak}} \cdot \frac{V_{2,\text{peak}}}{V_{1,\text{peak}}}$.

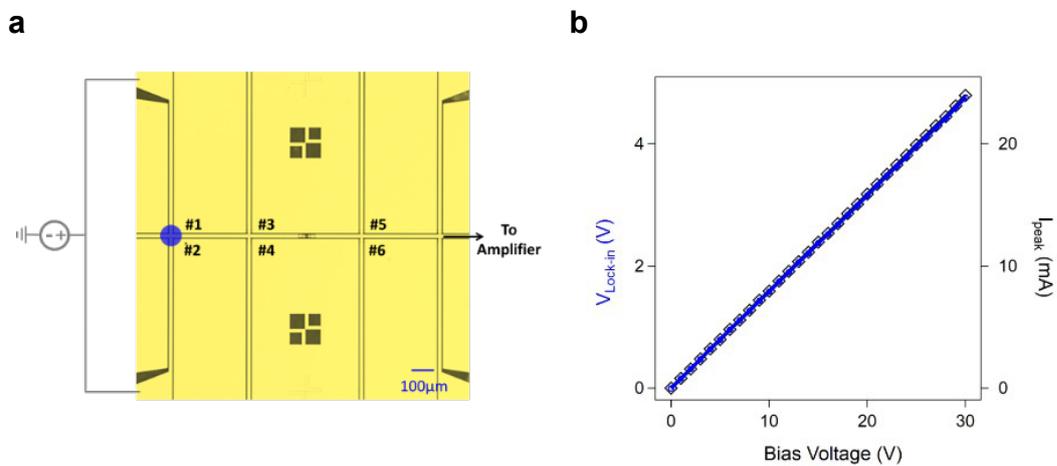

*Fig. S 13| Peak current value calibration for the launched electrical pulse. (a) Setup of measurements. (b) The signal from Lock-in amplifier (blue dot) and the calculated peak current (black square) versus bias voltage.*

## S3.7: Characterization of the $K_3C_{60}$ thin film

The surface morphology of the thin film was checked with atomic force microscopy before doping process. As shown in Fig. S 14, the thin film is polycrystalline with a domain size ~ 100 nm on average. The average domain thickness is estimated to be ~ 10 nm from the surface roughness.

A resistance versus temperature measurement is shown in Fig. S 15, where the resistance increases with increasing temperature, demonstrating the metallic behavior of the $K_3C_{60}$ thin film. This further emphasizes that the photo-induced state with reduced resistance is a non-thermal effect.

To confirm that the thin film was uniformly doped, the superconducting transition (resistance versus temperature) was measured with different bias currents as shown in Fig. S 16. The transition curves become broader with increasing bias current, but do not show a double transition feature, which has been observed as a typical characteristic in inhomogeneous superconductors [4-6]. This is indicative of homogeneous doping in our $K_3C_{60}$ thin film.

To further support the scenario that the $K_3C_{60}$ thin film undergoes a transition from a state that consists of weakly-coupled superconducting grains to a phase-coherent bulk superconductor when cooling through the broad superconducting transition, we plot $V$ versus $(I - I_c)$ with logarithmic scale in Fig. S 17, to get the value of $\alpha$ in the relation $V \propto (I - I_c)^\alpha$. The change of $\alpha$ from ~ 1 to 3 when the sample is cooled down from 15 K to 8 K, is consistent with this scenario as shown in previous studies [7, 8].

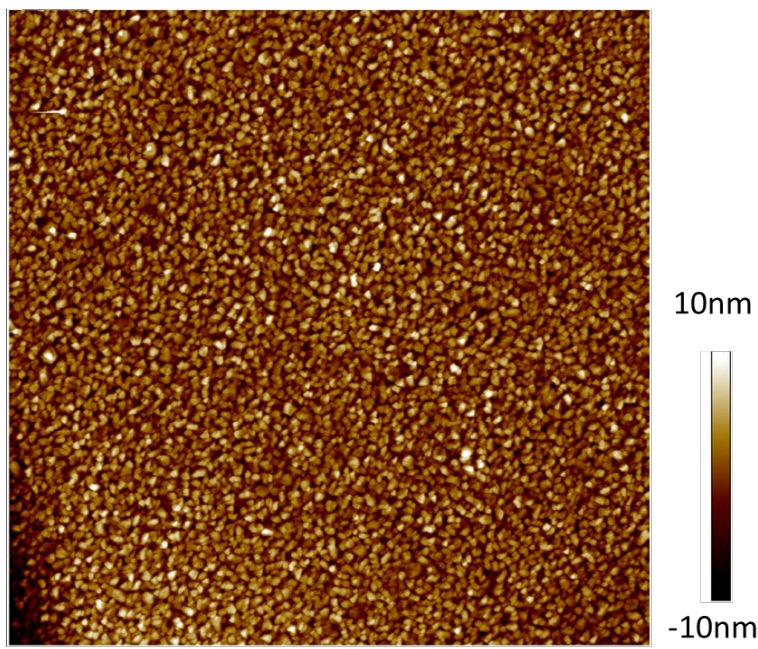

*Fig. S 14| A atomic force microscopy image of the $K_3C_{60}$ thin film. The image size is 5x5 μm.*

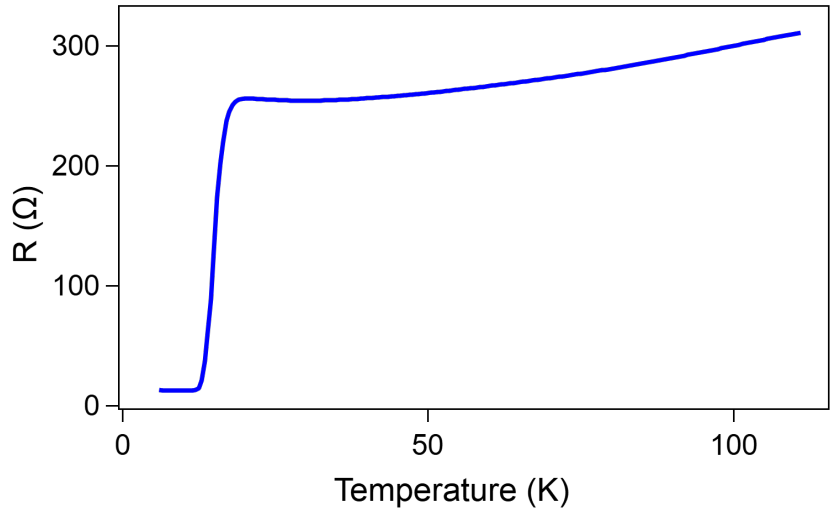

Fig. S 15| Resistance vs temperature measurement, with 1 µA bias current.

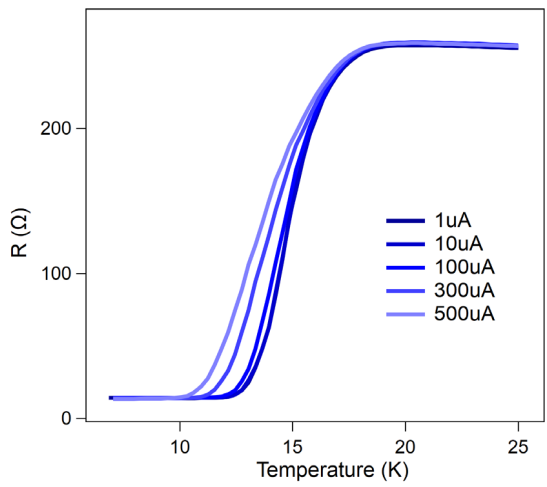

Fig. S 16| Resistance versus temperature measurements near $T_c$ with different bias currents.

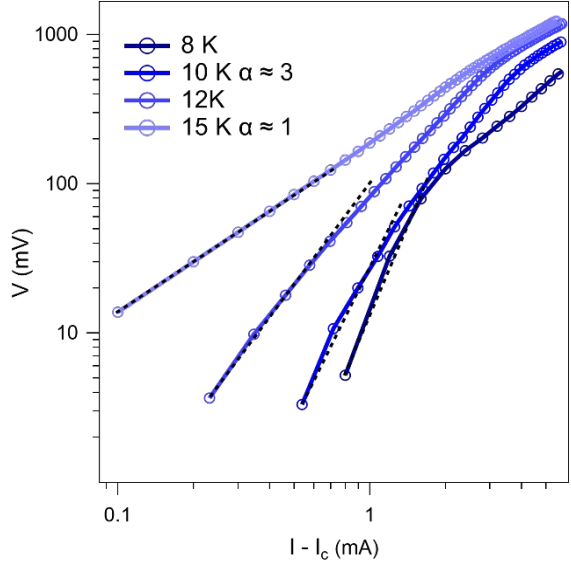

*Fig. S 17| Current-voltage characteristics of the $K_3C_{60}$ thin film below $T_c$.*

# S4. Data analysis

S4.1: Subtraction of the reflected signal

The raw data of the transmitted pulse $V_2(t)$ at 25 K is shown in Fig. S 18. At ~ 6 ps, there is a second peak in addition to the first peak at ~ 2 ps. While the first peak stems from the transmission of the incoming reference pulse $V_1(t)$, the second peak originates from the double reflection between switches #5,6 and the $K_3C_{60}$ thin film (indicated by the red arrows in Fig. S 18a), because the switches behave like a weak, local disturbance to the propagation of electrical pulse. The tail of the transmitted pulse is mixed with the rising edge of the second peak. To separate it from the measured data, we fit the rising edge with a Gaussian function (red curve in Fig. S 18b) and subtract it from the raw data.

This second peak also affects the measurements at other temperatures. Because the amplitude of this second peak is proportional to both the transmittance and reflectance of the sample, we calculate the rising edge at temperature T using the relation:

$$V_{\text{rise}}(t, T) = \frac{V_{2,\text{peak}}(T)(1 - V_{2,\text{peak}}(T))}{V_{2,\text{peak}}(25\ K)(1 - V_{2,\text{peak}}(25\ K))} \cdot V_{\text{rise}}(t, 25\ K)$$

where, $V_{\text{rise}}(t, 25\ K)$ is the fitted rising edge of the second peak at 25 K, and $V_{2,\text{peak}}(T)$ is the peak value of the transmitted pulse $V_2(t)$ (normalized by peak value of $V_1(t)$) at temperature T. There is a temperature dependence associated with the phase shift of the reflected pulse from the sample. To avoid this affecting our analysis, we only subtract the raw data with the rising edge up to a sampling time of 5 ps. The data after subtraction of the rising edge is shown in Fig. S 19.

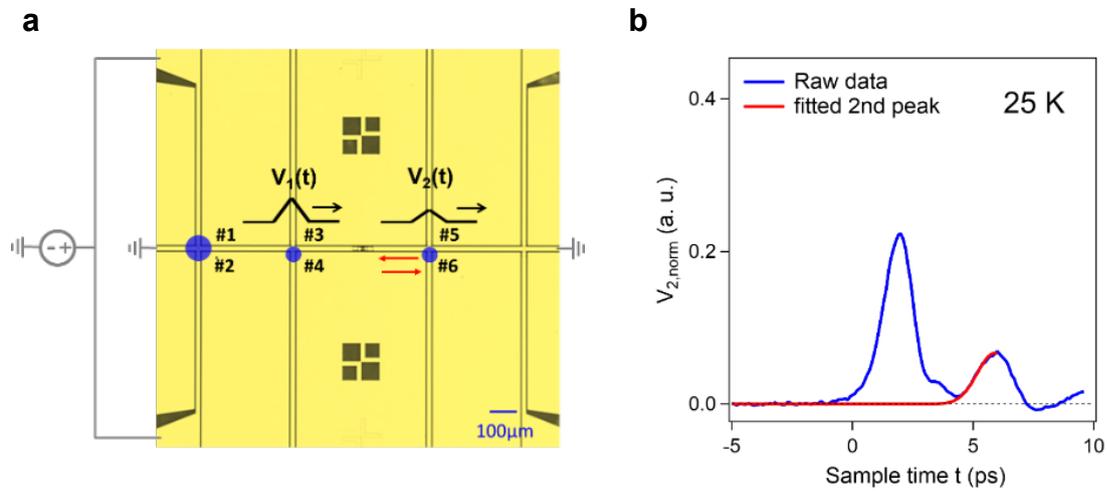

*Fig. S 18| Fitting of the second peak, which stems from the reflection between the photo-conductive switches and the sample.*

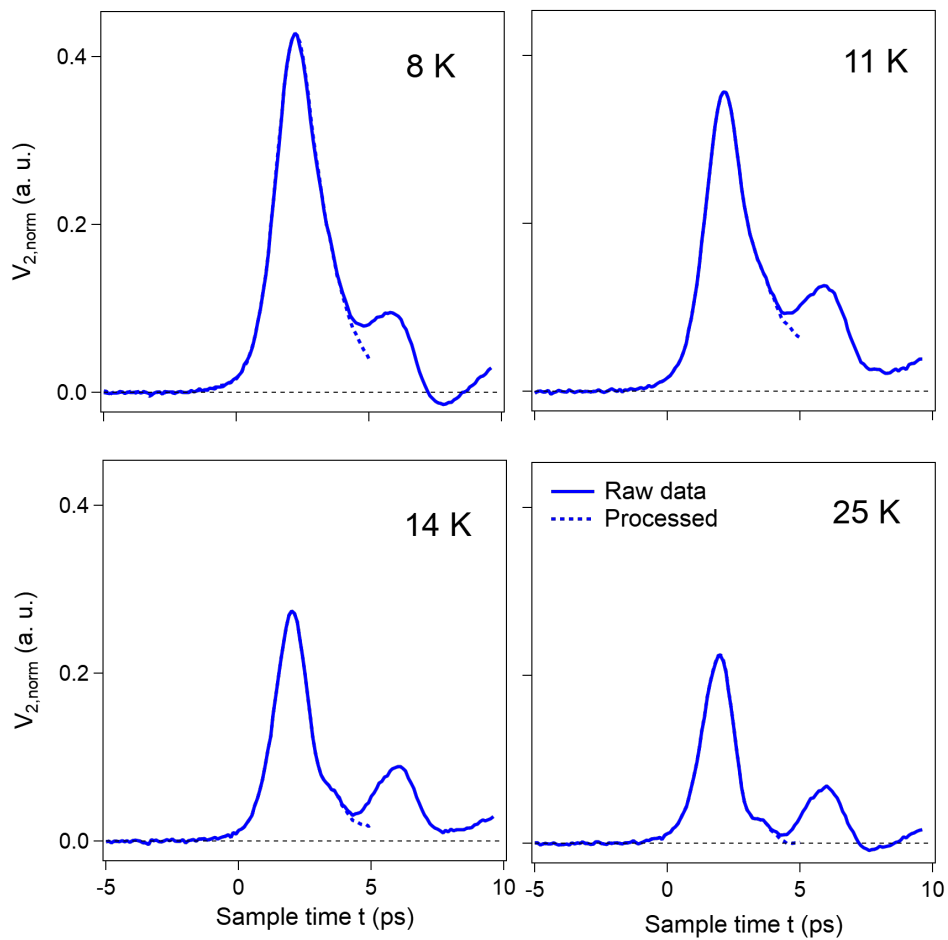

*Fig. S 19| Data after subtracting the rising edge of the second peak.*

## S4.2: Double-exponential fitting of data

For lower temperatures, the data, after subtracting the rising edge of the second peak, terminates abruptly at sampling time 5 ps. In order to extrapolate the pulse, we fit the tail of the data with a double-exponential function. As for the data in which the amplitude drops to zero at 5 ps, the data is extended by setting the value at later sampling times to zero. The quality of the fit and extrapolation of the data with a double-exponential function at lower temperatures is shown in Fig. S 20.

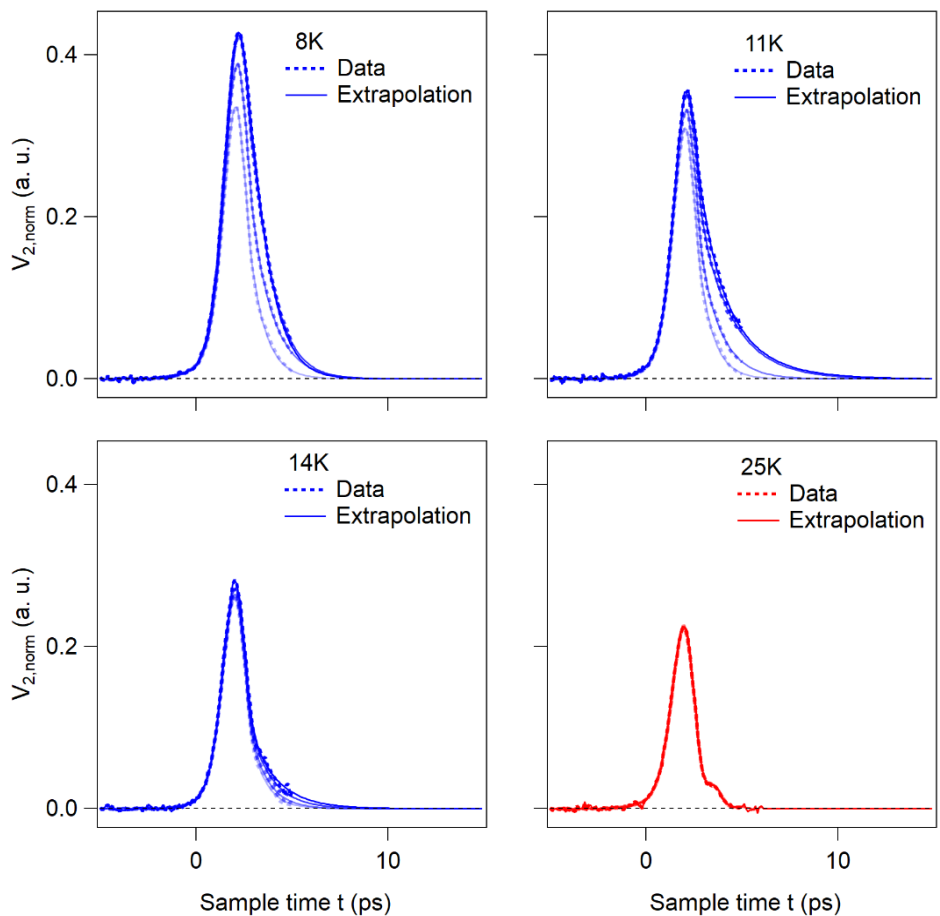

*Fig. S 20| Extrapolated data with double-exponential fitting.*

The reason why the fit and extrapolation shown in Fig. S 20 works well is as follows. In the electrical network shown below, $Z_w$ denotes the frequency-dependent impedance of the coplanar waveguide and $\hat{Z}$ the impedance of the sample. As shown in Section S3.5, the $Z_w$ can be treated as a pure resistance ~ 59 Ω. The reflectance $S_{11}$ and transmittance $S_{21}$ can be calculated from the relations below:

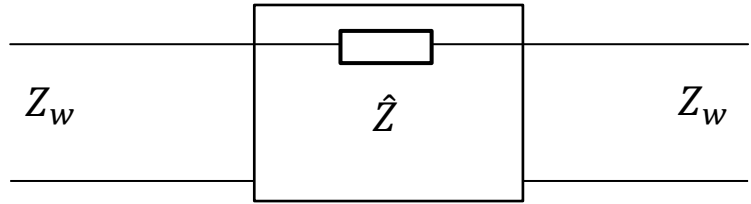

$$S_{11} = \frac{\hat{Z}}{2Z_w + \hat{Z}}$$

$$S_{21} = \frac{2Z_w}{2Z_w + \hat{Z}}$$

When sample is superconducting, we can model the sample with the following two-fluid circuit:

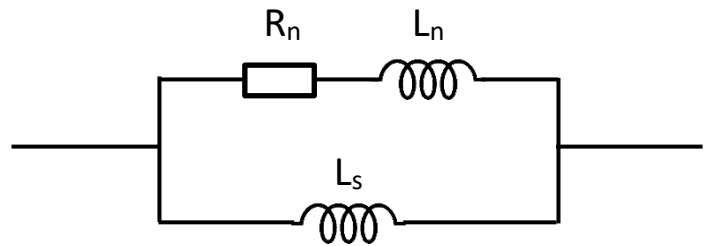

Its effective impedance is (using the Laplace transformation formalism):

$$\hat{Z}(s) = (R_n + sL_n) \parallel (sL_s)$$
$$= \frac{(R_n + sL_n) \cdot sL_s}{(R_n + sL_n) + (sL_s)}$$
$$= \frac{s \cdot R_n L_s + s^2 \cdot (L_n L_s)}{R_n + s \cdot (L_n + L_s)}$$

Therefore, the transmittance $S_{21}(s)$ is expressed as:

$$S_{21}(s) = \frac{2Z_w}{2Z_w + \hat{Z}(s)}$$
$$= \frac{2Z_w(L_n + L_s) \cdot s + 2Z_w R_n}{(L_n L_s) \cdot s^2 + [2Z_w(L_n + L_s) + R_n L_s] \cdot s + 2Z_w R_n}$$
$$= \frac{s + d}{b \cdot s^2 + c \cdot s + d}$$

Where,

$$b = \frac{L_n L_s}{2Z_w(L_n + L_s)}$$

$$c = 1 + \frac{R_n L_s}{2Z_w(L_n + L_s)}$$

$$d = \frac{R_n}{L_n + L_s}$$

To go back to the time-domain, the inverse-Laplace transformation yields:

$$\mathcal{L}_s^{-1}[S_{21}(s)](t) = a_1 \cdot e^{-t/\tau_1} + a_2 \cdot e^{-t/\tau_2}$$

Where,

$$1/\tau_1 = \left(\frac{c}{2b} + \frac{\sqrt{c^2 - 4bd}}{2b}\right)$$

$$1/\tau_2 = \left(\frac{c}{2b} - \frac{\sqrt{c^2 - 4bd}}{2b}\right)$$

$$a_1 = \frac{1}{\sqrt{c^2 - 4bd}} \left(\frac{\sqrt{c^2 - 4bd} + c}{2b} - 1\right)$$

$$a_2 = \frac{1}{\sqrt{c^2 - 4bd}} \left(\frac{\sqrt{c^2 - 4bd} - c}{2b} + 1\right)$$

The parameter values in the above equations are given by the relations: $R_n = \frac{R_{n0}}{n_n(T)}$, $L_n = \frac{L_{n0}}{n_n(T)}$, and $L_s = \frac{L_{s0}}{n_s(T)}$, where the Cooper pair density ratio $n_s(T) = 1 - \left(\frac{T}{T_c}\right)^4$, the normal carrier density ratio $n_n(T) = 1 - n_s(T)$, the resistance of normal state $R_{n0} = 250\ \Omega$ from experimental measurement, and the kinetic inductance of the normal state $L_{n0} = \tau \cdot R_{n0} = 55\ pH$ ($\tau$ from Ref. [9]). $L_{s0} = L_{n0} = 55\ pH$ is the kinetic inductance of the superconducting state at 0 K, assuming that the effective mass of a Cooper pair is double the mass of a normal carrier. At 8 K, from the calculation, we get $\tau_1 = 0.14\ ps$ and $\tau_2 = 0.52\ ps$, which are consistent with the fitting parameter values used for the data measured at 8 K.

## S5. More data

S5.1: Measurements with a nanosecond pulse generator

To bridge the gap between the nonlinear features of superconducting state observed with DC and picosecond current, we applied nanosecond voltage pulses to the $K_3C_{60}$ thin film from outside the chip, and monitored how the resistance of $K_3C_{60}$ changed at 8 K, with increasing the pulse duration and amplitude. The electronic setup is shown in Fig. S 21a, in which a pre-resistor (100 Ω) was placed before the sample in order to monitor the current flowing through the sample during the measurements. The voltages across the pre-resistor plus sample and across only the sample were monitored and recorded using an oscilloscope. Fig. S 21b and c show the voltage time-traces across the pre-resistor plus sample (red) and across only the sample (blue), for different durations of the applied voltage pulse. With a small voltage pulse bias (0.5 V), oscillations from the sample's inductive response are observed at early times and then the voltage becomes almost zero with a small contribution from the contact resistance. With a larger voltage pulse bias (6 V), the voltage across the sample starts to increase at early times and reaches equilibrium at later times in the 1 μs case. In the 150 ns case, the voltage pulse bias is gone before the voltage across the sample can stabilize.

The data points in the final 50 ns of the voltage pulse were used to calculate the bias current of the sample and subsequently the resistance. The resistance versus bias current curves measured at 8 K for different voltage pulse durations are summarized in Fig. S 22. As shown, for all pulse durations, the resistance starts to increase when the current increases above the same critical value. The change in resistance becomes steeper with longer pulse duration, which is consistent with what has been observed for DC and picosecond transport measurements.

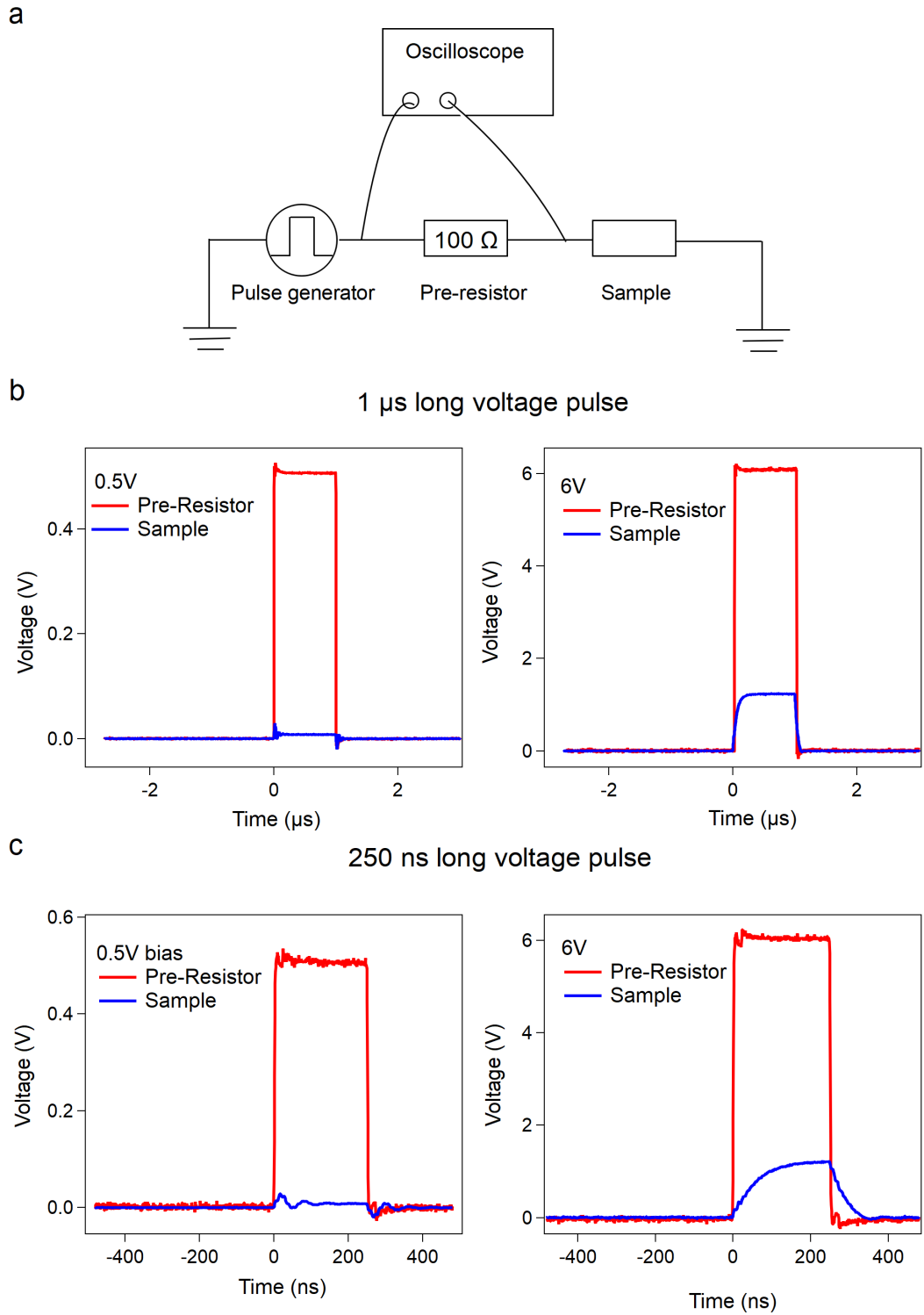

Fig. S 21| Time-trace of the voltage across the pre-resistor + sample (red) and across only the sample (blue) when the sample was biased using an external pulse generator. (a) Schematics of the measurement. The temperature of sample is kept at 8 K. (b) Two time-traces of the voltage potential for a 1 μs long voltage pulse with 0.5 V and 6V amplitude. (c) Same as (b) but for a 150 ns long voltage pulse.

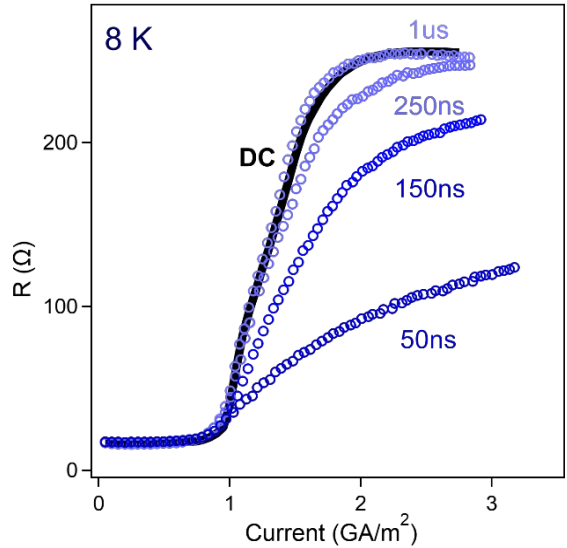

Fig. S 22| *Resistance versus bias current for different voltage pulse durations. The temperature of sample is kept at 8 K.*

## S5.2: Ultrafast transport measurements

### S5.2.1: Photo-induced state below and above $T_c$

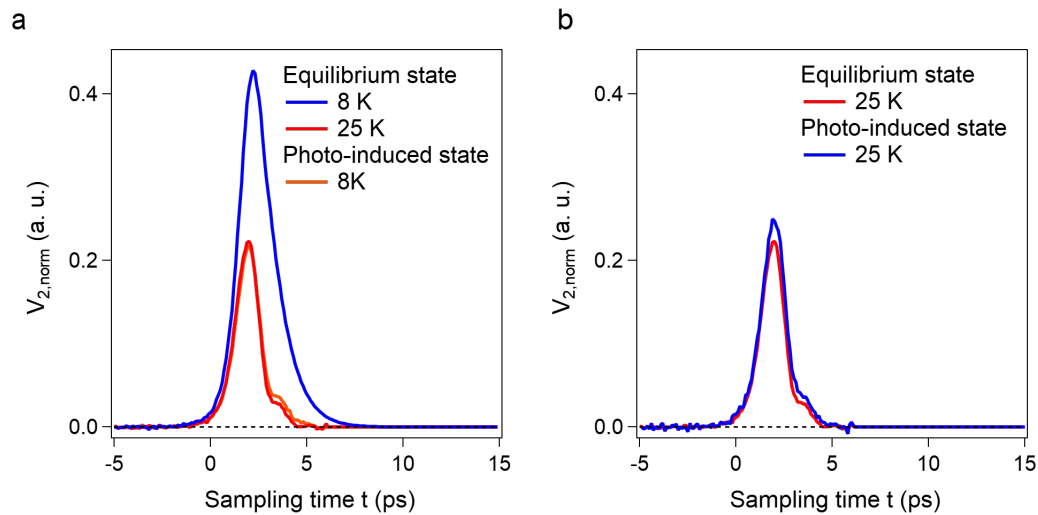

Fig. S 23| *Response of the photo-induced state at 8 K and 25 K. The fluence of the mid-IR pulses is 4mJ/cm$^2$ for both cases.*

Below $T_c$, the superconductivity is destroyed upon mid-IR excitation, as the photon energy is larger than the double of superconducting gap size. Although it is not yet fully understood why the effect of mid-IR excitation is different below and above $T_c$, the difference in properties of the initial states is a likely candidate.

The response of the photo-induced state at 8 K is shown in Fig. S 23a, which is similar to the response of the equilibrium state at 25 K. This can be understood, as the mid-IR excitation breaks the Cooper pairs into normal carriers. The response of the photo-induced state at 25 K is also shown (Fig. S 23b). Because of the formation of granular superconductivity compared with the bulk superconductivity at equilibrium 8 K, the transmittance is not as high as the equilibrium 8 K.

We also studied the lifetime of the photo-induced state at 8 K and 25 K. As shown in Fig. S 24, the lifetime of photo-induced state at 8 K is much longer than 2 ns, which is related with the dissipation of the energy injected by the mid-IR pulses into the $K_3C_{60}$ thin film. On the other hand, the lifetime of the photo-induced state at 25 K is around nanoseconds. This also makes the extraction of impedance of the photo-induced state at 50 GHz possible.

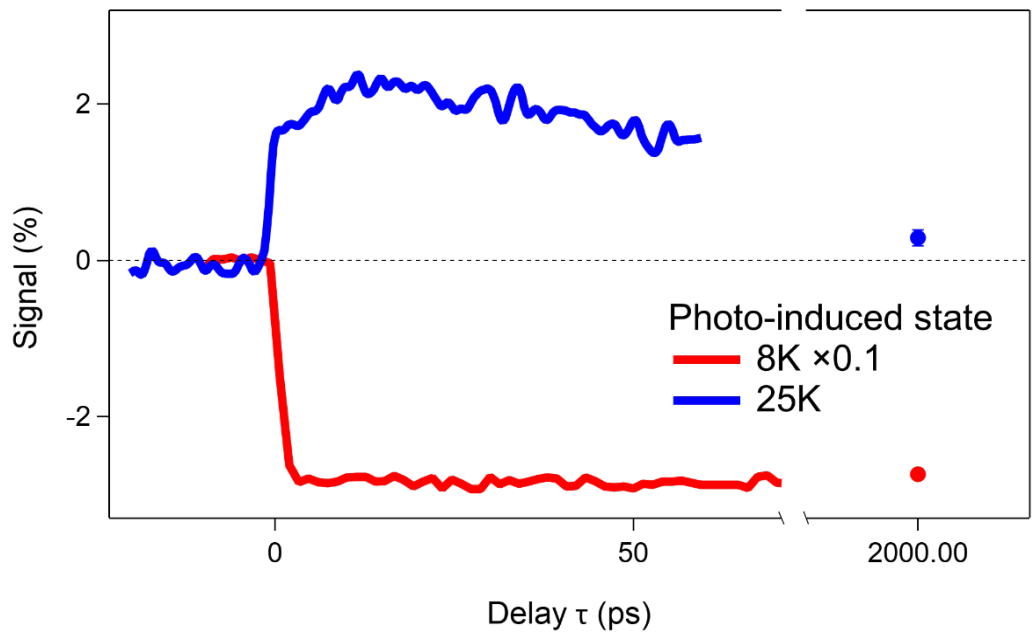

*Fig. S 24| Lifetime of the photo-induced state at 8 K and 25 K.*

S5.2.2: Fluence dependence

More fluence dependent data is shown in Fig. S 25. The transmittance at low frequency increases with stronger mid-IR pump and saturates at around 4 mJ/cm$^2$ (Fig. S 25a). The nonlinear behavior also gradually develops with increasing mid-IR pump fluence (Fig. S 25b).

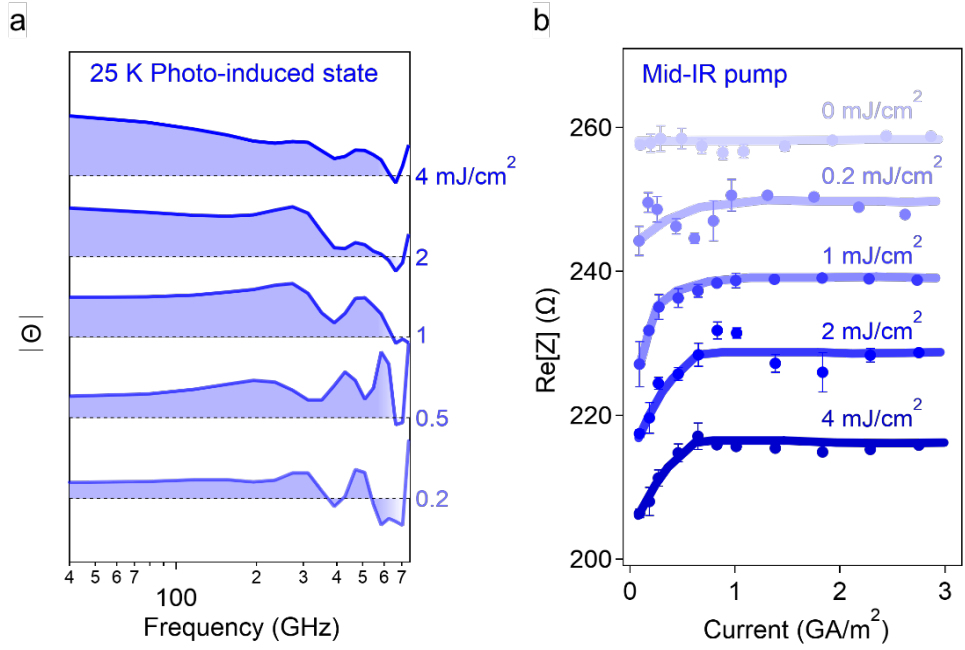

*Fig. S 25| Fluence dependent measurements of the photo-induced state. (a) Normalized transmittance $|\Theta|$ for different mid-IR fluences. The peak current density of the picosecond electrical pulse is 0.1 GA/m². The curves are shifted vertically for clarity and the dashed line indicates unity. (b) Calculated $Re[Z]$ at 50 GHz versus peak current density for different mid-IR pulse fluence.*

## S6. Simulation

S6.1: Time-domain simulation

The time-trace of the transmitted electrical pulse $V_2(t)$ was simulated with finite-element time-domain method using CST studio suite software, with the measured $V_1(t)$ and the optical conductivity of $K_3C_{60}$ extrapolated from data of Ref. [9] as input parameters. As shown in Fig. S 26, the data and simulation are alike.

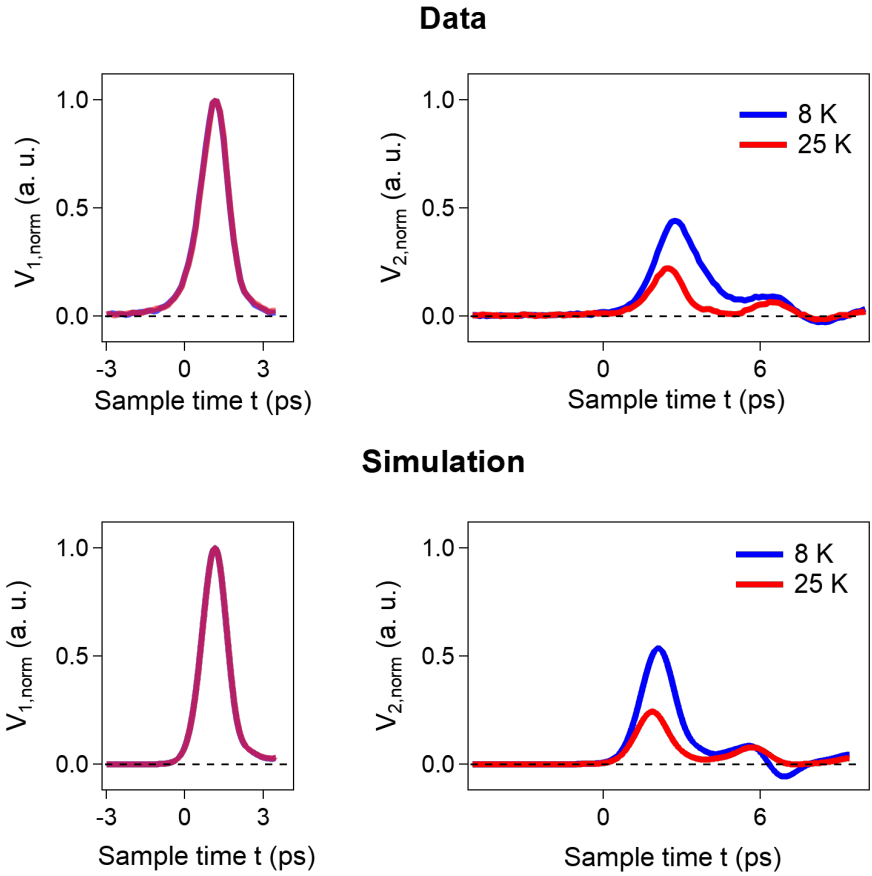

*Fig. S 26| Comparison of measured data and finite-element time-domain simulation of the transmitted electrical pulse.*

S6.2: Frequency-domain calculation

To explain the dip at frequency ~ 700 GHz observed in Fig. 2b, we calculated the normalized transmittance at 8 K in the frequency domain with Qucs Studio software by treating sample with two fluid circuit model used in Section S4.2. As shown in Fig. S 27, without capacitive coupling, the normalized transmittance of the superconducting state at 8 K is always larger than 1. This is contrasted with the case with a small capacitance contribution in parallel with the sample, in which the response could well reproduce the dip observed in the experiment. The value of the capacitance being used in the calculation is ~ 0.7 fF, which is congruous with the capacitance value between the two contacts of $K_3C_{60}$ film calculated with the method from Ref. [10].

To simulate the temperature dependence of $|\Theta|$, we need to consider the resistive response of the weak links between superconducting grains at intermediate temperatures. The detailed physical origin is discussed in Section S6.3. Here the measured resistance values of the sample (shown in Fig. 2 and Fig. S 16) are used as the values of the effective resistance $R_{WL}$, contributed by the weak links. The rest superconducting grains are treated with two-fluid circuit model. The calculated normalized transmittances at different temperatures are shown in Fig. S 28b, which could well reproduce the data shown in Fig. S 28c.

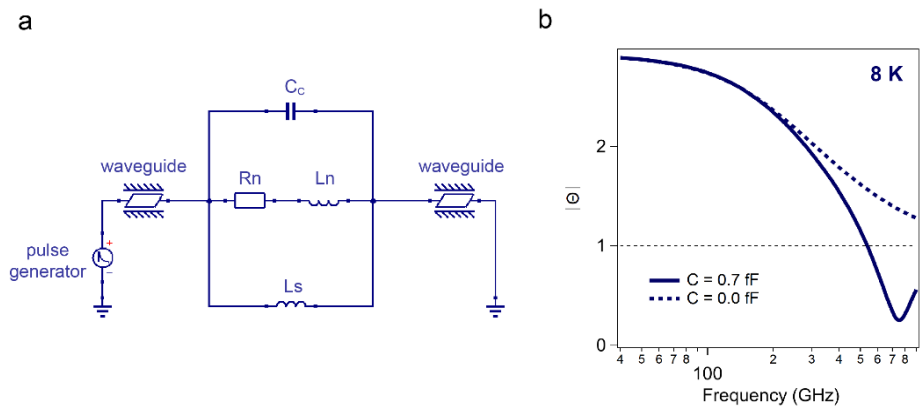

Fig. S 27| Transmittance calculation with Qucs Studio software.

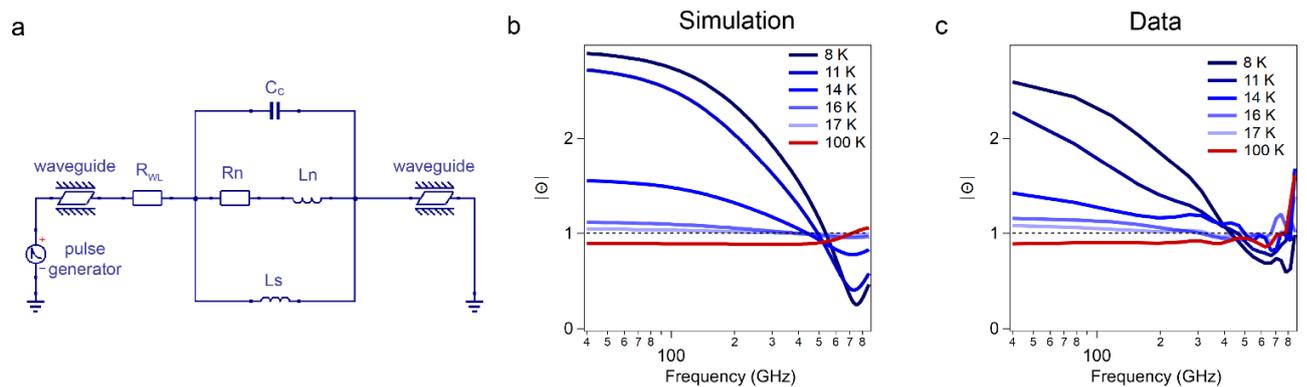

Fig. S 28| Transmittance calculation with Qucs Studio software, considering the resistive response of the weak links between the superconducting grains.

### S6.3: Critical current simulation

It has been known that the polycrystalline sample has a broader superconducting transition compared with a single-crystal sample, for the same superconducting material [8]. This has been attributed to the weak coupling between superconducting grains at intermediate temperatures [7, 8]. When the coupling energy between the superconducting grains ($\hbar I_c(T)/2e$, i.e., the energy cost for a phase slip) is comparable with the thermal fluctuation energy ($k_B T$), the phase slip can be thermally activated at weak links and the polycrystalline sample shows resistive response, despite the superconducting grains. Calculations can be done to check whether our $K_3C_{60}$ film falls into this range. From the AFM image shown in Fig. S 14, we estimate the average $K_3C_{60}$ grain size is 100 nm(W) x 100 nm(L) x 10 nm(T). We can therefore consider the 20 um(W) x 20 um(L) x 100 nm (T) sample to be a 200(W) x 200(L) x

10(T) mesh of grains, which are connected by weak links. A cross-section of this mesh contains 2000 weak links along the current direction. From the DC transport measurement shown in Fig. S 29a, we can make a first estimation of the low-temperature critical current $I_c(0\ K)$ of a single weak link of 3 mA / 2000 = 1.5 µA. Based on this, the coupling energy of a single weak link is ~ 2 meV at 10 K (given $I_c(T) = I_c(0\ K) * (1 - (T/T_c)^{1.5})$), which is comparable with the corresponding thermal fluctuation energy ~ 1 meV. This estimation shows that the competition between the coupling of the superconducting grains and the thermal fluctuation can affect the transport properties of the $K_3C_{60}$ thin film.

To simulate the effect of thermally activated phase slip on sample's transport properties, we describe the $V - I$ relation of a single weak link with the equations of motion for Josephson junction [11-13], which are shown below:

$$d\theta/dt = 2eV/\hbar \qquad (1)$$

$$C\ dV/dt = I - I_c(T)\sin\theta - \frac{V}{R} + \tilde{L}(t) \qquad (2)$$

Equation (1) is the Josephson condition relating $\theta$, the phase difference of the order parameter of two adjacent superconducting grains and V, the potential difference. Equation (2) describes the condition of conservation of charge. C is the capacitance of the junction; $I_c(T)$ is the maximum Josephson current at temperature T in the absence of noise; R is the resistance of the junction which is approximately the resistance in the normal state; $\tilde{L}(t)$ is the fluctuating noise current, which is thermal and given by $\langle \tilde{L}(t+\tau)\tilde{L}(t)\rangle = 2R^{-1}k_BT$. To get the $V - I$ relation of the sample, we consider that the sample has a 200(W) x 200(L) x 10(T) mesh of weak links as described before.

In the calculation, we estimate the value of C ~ $10^{-17} F$ from $\varepsilon_0 \cdot S/d$, where $\varepsilon_0$ is the vacuum electrical permittivity; S is the cross-section area of the weak link (200 x 10 nm²); d is the average distance between two adjacent superconducting grains, where we use 1 nm as a lower limit, because the domains are actually loosely touched as shown in the AFM image (Fig. S 14). Using smaller values of C does not change the calculation results, as the equations of motion already fall into the overdamped region as discussed in Refs. [11-13], which is also confirmed by our calculation using a smaller capacitance value. Other than the first estimation ~ 1.5 µA for $I_c(0\ K)$ of a single weak link, we need to use 3 µA to reproduce the measurements. This is because the phase-dependent supercurrent has been partially washed out by the thermal fluctuations [14]. The comparison between simulations and measurements for DC transport is shown in Fig. S 29. The experimentally observed features, including the finite resistance at small bias current and the vanishing critical current at intermediate temperatures, could be well reproduced. The steeper increase of the resistance in the measurements is presumably due to the heating effect, which becomes stronger when the sample gets resistive and the sample transits faster into the normal state, which is not included in the equations of motion for a weak link introduced above.

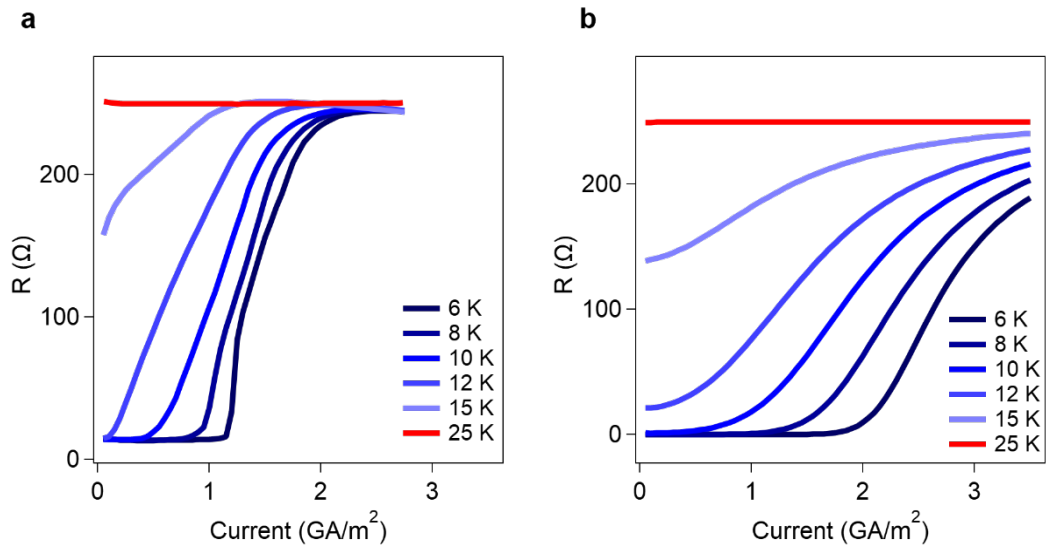

*Fig. S 29| Comparison between the data and calculations using equations of motion for Josephson junction.*

1. Kozich, V., A. Moguilevski, and K. Heyne, *High energy femtosecond OPA pumped by 1030nm Yb:KGW laser.* Optics Communications, 2012. **285**(21): p. 4515-4518.
2. Haddon, R.C., et al., *Electrical resistivity and stoichiometry of $K_xC_{60}$, $Rb_xC_{60}$, and $Cs_xC_{60}$ films.* Chemical Physics Letters, 1994. **218**(1): p. 100-106.
3. Hesper, R., et al., *Photoemission evidence of electronic stabilization of polar surfaces in $K_3C_{60}$.* Physical Review B, 2000. **62**(23): p. 16046-16055.
4. Gerber, A., et al., *Double-peak superconducting transition in granular L-M-Cu-O(L=Pr,Nd,Sm,Eu,D; M=Ce,Th) superconductors.* Physical Review Letters, 1990. **65**(25): p. 3201-3204.
5. Early, E.A., et al., *Double resistive superconducting transition in $Sm_{2-x}Ce_xCuO_{4-y}$.* Physical Review B, 1993. **47**(1): p. 433-441.
6. Jardim, R.F., et al., *Granular behavior in polycrystalline $Sm_{2-x}Ce_xCuO_{4-y}$ compounds.* Physical Review B, 1994. **50**(14): p. 10080-10087.
7. Abraham, D.W., et al., *Resistive transition in two-dimensional arrays of superconducting weak links.* Physical Review B, 1982. **26**(9): p. 5268-5271.
8. England, P., et al., *Granular superconductivity in $R_1Ba_2Cu_3O_{7-\delta}$ thin films.* Physical Review B, 1988. **38**(10): p. 7125-7128.
9. Degiorgi, L., et al., *Optical properties of the alkali-metal-doped superconducting fullerenes: $K_3C_{60}$ and $Rb_3C_{60}$.* Physical Review B, 1994. **49**(10): p. 7012-7025.
10. Naghed, M. and I. Wolff, *Equivalent capacitances of coplanar waveguide discontinuities and interdigitated capacitors using a three-dimensional finite difference method.* IEEE Transactions on Microwave Theory and Techniques, 1990. **38**(12): p. 1808-1815.

11. Ambegaokar, V. and B.I. Halperin, *Voltage Due to Thermal Noise in the dc Josephson Effect.* Physical Review Letters, 1969. **22**(25): p. 1364-1366.
12. Falco, C.M., W.H. Parker, and S.E. Trullinger, *Observation of a Phase-Modulated Quasiparticle Current in Superconducting Weak Links.* Physical Review Letters, 1973. **31**(15): p. 933-936.
13. Falco, C.M., et al., *Effect of thermal noise on current-voltage characteristics of Josephson junctions.* Physical Review B, 1974. **10**(5): p. 1865-1873.
14. Tinkham, M., *Introduction to Superconductivity*. 2004: Dover Publications.